\documentstyle[epsf,graphics,general_cite]{mn}
\newif\ifAMStwofonts

\def\mc{\multicolumn}

\ifoldfss
  \ifCUPmtlplainloaded \else
    \NewTextAlphabet{textbfit} {cmbxti10} {}
    \NewTextAlphabet{textbfss} {cmssbx10} {}
    \NewMathAlphabet{mathbfit} {cmbxti10} {} 
    \NewMathAlphabet{mathbfss} {cmssbx10} {} 
  \fi
  \ifAMStwofonts
    \ifCUPmtlplainloaded \else
      \NewSymbolFont{upmath} {eurm10}
      \NewSymbolFont{AMSa} {msam10}
      \NewMathSymbol{\upi}     {0}{upmath}{19}
      \NewMathSymbol{\umu}     {0}{upmath}{16}
      \NewMathSymbol{\upartial}{0}{upmath}{40}
      \NewMathSymbol{\leqslant}{3}{AMSa}{36}
      \NewMathSymbol{\geqslant}{3}{AMSa}{3E}

       \let\le=\leqslant
       \let\ge=\geqslant
    \fi
  \fi
\fi 

\ifnfssone
  \newmathalphabet{\mathit}
  \addtoversion{normal}{\mathit}{cmr}{m}{it}
  \addtoversion{bold}{\mathit}{cmr}{bx}{it}
  \newmathalphabet{\mathbfit} 
  \addtoversion{normal}{\mathbfit}{cmr}{bx}{it}
  \addtoversion{bold}{\mathbfit}{cmr}{bx}{it}
  \newmathalphabet{\mathbfss} 
  \addtoversion{normal}{\mathbfss}{cmss}{bx}{n}
  \addtoversion{bold}{\mathbfss}{cmss}{bx}{n}
  \ifAMStwofonts
    \ifCUPmtlplainloaded \else
      \UseAMStwoboldmath
      \makeatletter
      \new@mathgroup\upmath@group
      \define@mathgroup\mv@normal\upmath@group{eur}{m}{n}
      \define@mathgroup\mv@bold\upmath@group{eur}{b}{n}
      \edef\UPM{\hexnumber\upmath@group}
      \new@mathgroup\amsa@group
      \define@mathgroup\mv@normal\amsa@group{msa}{m}{n}
      \define@mathgroup\mv@bold\amsa@group{msa}{m}{n}
      \edef\AMSa{\hexnumber\amsa@group}
      \makeatother
      \mathchardef\upi="0\UPM19
      \mathchardef\umu="0\UPM16
      \mathchardef\upartial="0\UPM40
      \mathchardef\leqslant="3\AMSa36
      \mathchardef\geqslant="3\AMSa3E

       \let\le=\leqslant
       \let\ge=\geqslant
    \fi
  \fi
\fi 

\ifnfsstwo
  \DeclareMathAlphabet{\mathbfit}{OT1}{cmr}{bx}{it}
  \SetMathAlphabet\mathbfit{bold}{OT1}{cmr}{bx}{it}
  \DeclareMathAlphabet{\mathbfss}{OT1}{cmss}{bx}{n}
  \SetMathAlphabet\mathbfss{bold}{OT1}{cmss}{bx}{n}
  \ifAMStwofonts
    \ifCUPmtlplainloaded \else
      \DeclareSymbolFont{UPM}{U}{eur}{m}{n}
      \SetSymbolFont{UPM}{bold}{U}{eur}{b}{n}
      \DeclareSymbolFont{AMSa}{U}{msa}{m}{n}
      \DeclareMathSymbol{\upi}{0}{UPM}{"19}
      \DeclareMathSymbol{\umu}{0}{UPM}{"16}
      \DeclareMathSymbol{\upartial}{0}{UPM}{"40}
      \DeclareMathSymbol{\leqslant}{3}{AMSa}{"36}
      \DeclareMathSymbol{\geqslant}{3}{AMSa}{"3E}

       \let\le=\leqslant
       \let\ge=\geqslant
    \fi
  \fi
\fi 

\ifCUPmtlplainloaded \else
  \ifAMStwofonts \else 
    \def\upi{\pi}
    \def\umu{\mu}
    \def\upartial{\partial}
  \fi
\fi

\title[Imaging H$_{2}$O clouds around 4 AGB stars]{Sub-au imaging of water vapour clouds around four Asymptotic Giant Branch stars}
\author[Bains et al.]
       {I.~Bains,$^{1,2}$ R.J.~Cohen,$^2$ A.~Louridas,$^{3,2}$
A.M.S.~Richards,$^2$ 
 \newauthor
D. Rosa-Gonz\'alez$\;^{4,2}$ and J.A.Yates$\;^{5,1}$
 \\ 
$^1$Astronomy Group, Department of Physical Sciences, University of
        Hertfordshire, College Lane, Hatfield, Hertfordshire, AL10 9AB, UK.\\ 
$^2$Jodrell Bank Observatory, University of Manchester, Macclesfield,
        Cheshire, SK11 9DL, UK.\\
$^3$Electronic Engineering Laboratory, University of Kent at Canterbury, Canterbury, Kent CT2 7NT, UK.\\
$^4$INAOE, Luis Enrrique Erro 1,Tonantzintla, Puebla
        72840. M\'exico. \\
$^5$Department of Physics and Astronomy, University College London,
                  Gower Street, London, WC1E 6BT, UK.\\
}
\date{ }

\pagerange{\pageref{firstpage}--\pageref{lastpage}}
\pubyear{}
\begin{document}
\bibliographystyle{mnras}

\maketitle

\label{firstpage}


\begin{abstract}

We present MERLIN maps of the 22-GHz H$_{2}$O masers around four 
low-mass late-type stars (IK~Tau U~Ori, RT~Vir and U~Her), made 
with an angular resolution of $\sim$15 milliarcsec  
and a velocity resolution of 0.1 km s$^{-1}$.  
The H$_{2}$O masers are found 
in thick expanding shells with inner radii $\sim$6 to 16~au and 
outer radii four times larger.  
The expansion velocity increases radially through the H$_{2}$O maser regions, 
with logarithmic velocity gradients of 0.5--0.9.  
IK Tau and RT Vir have well-filled H$_{2}$O maser shells 
with a  spatial offset between the near and far sides of the shell, 
which suggests that the masers are distributed in oblate
spheroids inclined to the line of sight.
U Ori and U Her have elongated 
poorly-filled shells with indications that the
masers at the inner edge have been compressed by shocks; these stars
also show OH maser flares.
MERLIN resolves individual maser clouds, which have 
diameters of 2 -- 4 au and filling factors of only $\sim$0.01 with respect to the whole H$_{2}$O maser shells.  
The CSE velocity structure gives additional evidence the maser 
clouds are density bounded.  
Masing clouds can be identified over a similar timescale to their sound
crossing time ($\sim2$ yr) but not longer.  
The sizes
and observed lifetimes of these clouds are an order of magnitude smaller
than those around red supergiants, similar to the ratio of low-mass:high-mass stellar masses and sizes.  This 
suggests that cloud size is determined by stellar properties, not local
physical phenomena in the wind.  

\end{abstract}

\footnotetext{Contact e-mail: ib@star.herts.ac.uk, amsr@jb.man.ac.uk}

\section{Introduction}
\label{intro}

\begin{table*}
\caption{The properties of the stars and their circumstellar envelopes.} 
\label{sources}
\begin{tabular}{lcllcllrl}
Source  &\mc{1}{c}{Position (J2000)}&$V_*$&$P$&$m_{\rm V}$&$d$    &$\dot{M}_{\rm CO}$ &$V_{\rm max}$ & $r_{\rm CO}$\\
	&(hh mm ss.sss+dd mm ss.ss)& (km s$^{-1}$)             &(days)
	&\mc{1}{c}{(magnitude)}  &(pc)      &  M$_{\odot}$ yr$^{-1}$ & (km s$^{-1}$)&  (au)   \\  
(1)	&\mc{1}{c}{(2)}     &(3)	   &(4)   &\mc{1}{c}{(5)}            &(6)&(7)&(8)&(9)	\\
\hline	  	 		  	 		        		  		 		    			 
IK Tau	&03 53 28.83$\;$ +11 24 20.5$\;$     	&$+34.0$	&470&10.8 -- 16.5&$265\pm^5_{45}$ & 2.6$\times10^{-6}$&21.3&6320, 26000	\\
U Ori   &05 55 49.169  +20 10 30.69		&$-39.5$	&368&4.8 -- 13.0&$260\pm50$	 &2.3$\times10^{-7}$&7.5&7000	\\
RT Vir &13 02 37.982  +05 11 08.38		&$+18.2$   &155,&7.4 -- 8.7&$133\pm20$&1.3$\times10^{-7}$&8.8& ($2035\times1100$)	\\
&&&\mc{2}{l}{$112\pm11$}&&&&$\pm320$ p.a. 30\degr\\
&&& \mc{2}{l}{\& $170\pm7$} &&&&\\
U Her   &16 25 47.471  +18 53 32.87		&$-14.5$  &406&6.4 -- 13.4&$190\pm70$, 	 &3.4$\times10^{-7}$&11.5&5400	\\
&&&&& $385\pm5$ 	&&&\\
\hline
& \mc{7}{c}{References}\\
IK Tau & \mc{1}{c}{M98}      &K87 & G98&G98& H97& O98, B00&&B89, K85\\
U Ori& \mc{1}{c}{H97}	&C91& G98&G98& C91& K98&&Y95\\	
RT Vir& \mc{1}{c}{H97} &N86& G98, E01&G98& H97& K98, K99&&K95\\
U Her& \mc{1}{c}{H97} &C94 & G98 &G98& vL00, C94& Y95&&Y95\\
\hline
\end{tabular}
\flushleft{NOTES:\\
(2) The position of IK Tau is accurate to better than 1\arcsec, 
the positions of the other stars are accurate to $\sim10$~mas.\\
(3) $V_{\rm LSR}$ measurements derived from OH and SiO 
maser data, accurate to better than 0.5~km~s$^{-1}$. \\
(4) RT Vir varies irregularly. OH data (E01) suggests 
multiple periodicity.  The periods of the other stars 
are accurate to within 10 days.\\
(5) IK Tau and RT Vir are Zodiacal and have poorly sampled optical periods; IK Tau is also heavily obscured by its thick CSE.\\
(6) 
In this analysis we adopt a distance of 133 pc for RT Vir and 
266 pc for the other sources.\\
(7) We used the distances given in note (6) and the most recent CO data available to
calculate $\dot{M}_{\rm CO}$ by the method of \protect{\scite{Groenewegen99}}. Almost all available CO and
IR data from the  literature give results which agree to within a factor of 5.  \\
(8) Maximum CO or thermal SiO expansion velocity, references as for (6). The discrepancies between these
values and previous measurements is $\approx{^{+1}_{-3}}$ km s$^{-1}$.\\
(9) Maximum radius of CO emission, adjusted to our adopted distances.  
\\
REFERENCES:  \\
B00 \scite{Bieging00};
B89 \scite{Bujarrabal89};
C91 \scite{Chapman91};
C94 \scite{Chapman94};
E01 \scite{Etoka01};
G98 \scite{Kholopov98};
H97 \scite{ESA97};
K85 \scite{Knapp85}
K87 \scite{Kirrane87};
K98 \scite{Knapp98};
K99 \scite{Kerschbaum99};
M98 \scite{Monet98};
N86 \scite{Nyman86};
O98 \scite{Olofsson98};
S95 \scite{Stanek95};
Y95 \scite{Young95};
vL00 \scite{vanLangevelde00}.}
\end{table*}

Stars which start life with a mass $M_*$ of one or a few M$_{\odot}$ usually
end up as 0.6 -- 1.0 M$_{\odot}$  white dwarves (WDs). Most of their mass loss
occurs during and at the end of the Asymptotic Giant Branch (AGB)
stage, at rates of $\dot{M}\approx$~($10^{-8}$ -- $10^{-4}$) M$_{\odot}$ yr$^{-1}$,
resulting in dusty, thick circumstellar envelopes (CSEs).  Mass loss
on the AGB determines the final stages of evolution of low- and
intermediate-mass stars and significantly contributes to the chemical
evolution of galaxies, in particular providing up to 80 percent of the
dust \cite{Whittet92}.  

AGB stars contain a degenerate WD-like core surrounded by a shell
enriched with CNO-cycle elements and other products of
high-temperature nucleosynthesis, which are brought to the stellar
surface by convection and major dredge-up events known as thermal
pulses.  As stars enter the AGB their optical pulsation periods $P$
become longer and more regular, of the order of a year. The outer
layers are cool (typically 2000 -- 2500 K) and very extended (radius
$R_* \ga 1$ au), with a luminosity $L_*$ of a few $10^3$
L$_{\odot}$. The CSE becomes thicker during the AGB lifetime of a few
times $10^4 - 10^5$ yr \cite{Jura93}.  More massive stars have shorter
lifetimes but lose mass at a higher rate (Vassiliadis \& Wood 1993,
1994; \nocite{Vassiliadis93} \nocite{Vassiliadis94}
\pcite{Blocker95}).  Finally, the remnants of the stellar shell heat up
and are lost in a superwind, leading to a planetary nebula (PN)
surrounding a WD.

 \scite{vanderVeen88} located Miras, OH/IR stars and proto-PNe (PPNe)
on the IRAS colour-colour diagram along a curve corresponding to
increasingly thicker and cooler CSEs.  \scite{Jura93} suggest Mira
periods continue to lengthen on the AGB but \scite{Whitelock94}
present evidence that Miras with $P\la400$ days are a separate
population from longer-period objects. Semi-regular variables (SRa,b)
are generally reported to have lower mass loss rates than Miras and
could be stars entering \cite{Kerschbaum92} or leaving \cite{Young93}
the AGB.

It is generally agreed that the stellar pulsations levitate the
photosphere \cite{Bowen88} until it is cool enough for dust grains to
nucleate and grow at a few $R_*$, as described by Gail \& Sedlmayr
(1998a, 1998b, 1999). \nocite{Gail98a} \nocite{Gail98b}
\nocite{Gail99}  Radiation pressure on dust then drives the wind away
from the star via collisions with the gas.

The CSEs of O-rich stars produce masers including SiO, H$_2$O 22-GHz,
OH mainline and OH 1612-MHz emission \cite{Cohen89}.  High-resolution
mapping of the winds from red supergiants (RSG) of $M_*\ga$ 10 M$_{\odot}$ shows discrete H$_{2}$O vapour clouds $5
- 15$ au in radius which are $1 - 2$ orders of
magnitude denser than the surrounding wind \cite{Richards99}.  No
direct measurements of the unbeamed sizes of H$_{2}$O maser clouds
around Miras have previously been published. 
The H$_2$O maser regions of AGB stars at distances of a few
100~pc are $\sim100$ mas in radius. 

We observed four nearby AGB stars, IK Tau, U Ori, RT Vir and U Her, at
 22~GHz using MERLIN.  Its baselines correspond to angular scales from
 about 10 to 200 milli-arcseconds (mas), allowing us to image H$_2$O
 masers at high resolution and still detect all of the extended flux (Section~\ref{maps}).  All
 four stars have 9.7-$\mu$m silicate features in emission \cite{olnon86} and similar warm IRAS colours, $\log(F_{25}/F_{12})<0$, where $F_{25}$ and $F_{12}$ are the 25 and 12 $\mu$m fluxes respectively.  Other properties taken
 from the literature are given in Table~\ref{sources}.  The absolute
 position of the star is given in column (2) and its velocity $V_*$
 with respect to the local standard of rest ($V_{\rm LSR}$) is given
 in column (3).  The period $P$ is given in column (4). IK Tau, U Ori
 and U Her have regular periods and are classed as Mira variables; RT
 Vir is an SRb with no clear optical period. The optical magnitude
 range $m_{\rm V}$ is given in column (5) of Table~\ref{sources}.  The distance $d$ in column
 (6) is most reliable for RT Vir.  The estimates for IK Tau and U Ori
 using various methods agree with the results of the period-luminosity
 relationship within the uncertainties.  For U Her this method gives
 double the distance derived from OH parallax.  For all three Miras,
 the values of $d$ are model dependent and for simplicity in all
 future calculations we use the {\em Hipparcos} distance of 133 pc for
 RT Vir and 266 pc for the other stars.  The values quoted for the
 mass loss rates $\dot{M}_{\rm CO}$ in column (7) assume a gas-dust
 ratio of $120 - 200$ and a fractional CO number density of $(2 -
 5)\times10^{-4}$.  The CO data also provide the maximum expansion
 velocities and sizes of the CSEs in columns (8) and (9) respectively,
 see Section~\ref{shells}.
In Section~\ref{obs+red} we describe our observations and
in Section~\ref{results} we present the observational results.
Our analysis of the kinematics and dynamics of the circumstellar 
envelopes is given in Section~\ref{discuss}.
We investigate the properties of individual water maser
clouds in Section~\ref{analysis}. 
We summarise our findings on maser clouds and
the properties of AGB stellar winds in Section~\ref{conclusions}.

\section{Observations and data reduction. }
\label{obs+red}

\begin{table}
\caption{Details of MERLIN 22-GHz observations, including the
restoring beams and noise levels in quiet channels for datacubes made
at full resolution.}
\begin{tabular}{llllrr}
\hline
Source	& Date&Integ&$V_{\rm LSR}$&$\Theta_{\rm B}$	&$\sigma_{\rm rms}$\\
	&(yymmdd)& time	 & (km s$^{-1}$) &(mas)&\mc{1}{l}{(mJy}  	\\
& & (hr) &&&bm$^{-1}$)\\
(1)	&   (2)	&(3)	&(4)		&(5)&(6)			\\
\hline	                                                                                     
IK Tau  &940415 &11.0  &$+34$	  	&15	&10\\
U Ori   &940417 &12.6  &$-39$		&15	&12\\
RT Vir  &940416 &11.8  &$+18$	  	&20	&12\\
U Her   &940413 &13.7  &$-15$	 	&15	&14\\
\hline
\end{tabular}
\label{obs}
\end{table}

The four stars were observed in 1994 April using 5 telescopes of MERLIN:  
the Mk2 telescope at Jodrell Bank and outstation telescopes at 
Pickmere, Darnhall, Knockin and Cambridge.  The 
maximum baseline was 217 km, giving a minimum fringe spacing of 
13~mas at 22.235~GHz.  
Table~\ref{obs} gives further details of the observations, 
including the date and duration of the observations and the 
central velocity of the MERLIN spectral band $V_{\rm LSR}$ (columns (2),  
(3) and (4) respectively).  
Here and elsewhere radial velocities are given with respect to 
the Local Standard of Rest (LSR).  
We used a spectral 
bandwidth of 2~MHz divided into 256 frequency channels, which gave a 
velocity resolution of 0.1~km~s$^{-1}$.  
The quasar
3C84 was observed for $\sim$1 hr once or twice at the appropriate
frequency for each source.  It had a flux density of $23\pm1$ Jy at
that time (Ter\"{a}sranta, priv. comm.)  and was used to calibrate the
bandpass and set the flux scale for each source, with a final accuracy
of $\sim10$ percent.  

We reduced the data as described in \scite{Richards99} for 22-GHz
observations, using local MERLIN-specific programs and {\sc aips}.
LHC and RHC polarisation data were observed and calibrated separately
and simultaneously, but all maps were made in total intensity. About
10 percent of the data were unusable, mainly due to bad weather. We
derived corrections for phase and amplitude errors using
self-calibration only, so we could not obtain accurate absolute
positions. We reweighted the calibrated data to attain the optimum
combination of resolution and sensitivity and mapped and {\sc clean}ed
each source using the {\sc aips} task {\sc imagr}.

The FWHM (full width half maximum) of the restoring beam ($\Theta_{\rm
B}$) is given in column (5). The MERLIN beam at these declinations is
moderately elliptical but using a circular restoring beam of
equivalent area did not introduce any artefacts and makes the maps
easier to interpret.  The typical rms noise for quiet
channels,  $\sigma_{\rm rms}$,  is given in column (6).
In the presence of bright emission (exceeding 1~Jy~beam$^{-1}$) 
the noise can be $\sim1$ percent of the channel peak due to
deconvolution errors arising from the sparse baseline coverage.
Note that these parameters refer to the data cubes mapped at full
velocity resolution (0.105 km s$^{-1}$), which were used for all
quantitative analysis of individual maser clouds.  
Additional maps were made at 
lower spatial and velocity resolution to study the overall 
properties of the H$_{2}$O envelopes at 
greater sensitivity.  Some of these maps are presented in
Figs.~\ref{IKTAU} to~\ref{RTVIR}.

We fitted 2-D elliptical Gaussian components to each patch of emission
in each channel in each datacube in order to measure the peak flux
density $I_{\rm c}$, the position relative to the reference feature
used for self-calibration, the total area and the total flux density
$S_{\rm c}$.  The relative position uncertainty $\sigma_{\rm pos}$ is
proportional to the (dirty beam size)/(signal to noise ratio), as
described analytically in \scite{Condon97} and \scite{Condon98} and
adapted for the MERLIN beam by \scite{Richards97t} and
\scite{Richards99}.  This is typically 1~mas for an isolated component
with $I_{\rm c}\approx1$~Jy~bm$^{-1}$ and 0.1~mas for $I_{\rm
c}\ga10$~Jy~bm$^{-1}$. The FWHM of each component, $s$, was found by
deconvolving the restoring beam from the component area, with an
uncertainty $\sigma_{\rm s}=\sqrt2\sigma_{\rm pos}$.

The fitted components were grouped into features if three or more
components with $I_{\rm c}>3\sigma_{\rm rms}$ occurred in adjacent channels
with positions overlapping to within the position error or component
size.  Non-matched components were discarded, as were any others which
coincided with the beam sidelobe structure.  

\section{Observational Results}
\label{results}

\begin{table*}
\caption{Global properties of the H$_{2}$O maser envelopes.}
\begin{tabular}{lrcrrrrrr}
\hline
Source	& $\Delta\theta_{\rm tot}$ & V$_{*}$ & \mc{2}{c}{$V_{\rm LSR}$ range}& $\Delta$$V_{\rm LSR}$&$S_{\rm peak}$&$S_{\rm tot}$ \\
 & (mas) & (km~s$^{-1}$) &min & max&\mc{1}{l}{(km~s$^{-1}$)} &(Jy)&\mc{1}{l}{(Jy km~s$^{-1}$)}\\
 & & & \mc{2}{c}{(km s$^{-1}$)} & & & \\
(1)	&   (2)	&(3)	&(4)		&(5) &(6)&(7) & (8)	\\
\hline	                                                                                     
IK Tau &445 & +34.0 &$+21.2$ & $\ge+46.7$	 &$\ge25.5$ &25& $\ge$108\\
U Ori  &235 & -39.5 &$-41.5$ & $-34.8$ 	 &6.7 &26&20 \\	
RT Vir &370 & +18.2 &$+9.5$ & $+26.6$      	&17.1&394&755 	\\
U Her  &300 & -14.5 &$-22.4$ & $-9.8$	&12.6&22&45  \\
\hline
\end{tabular}
\label{peaks}
\end{table*}

\begin{figure*}
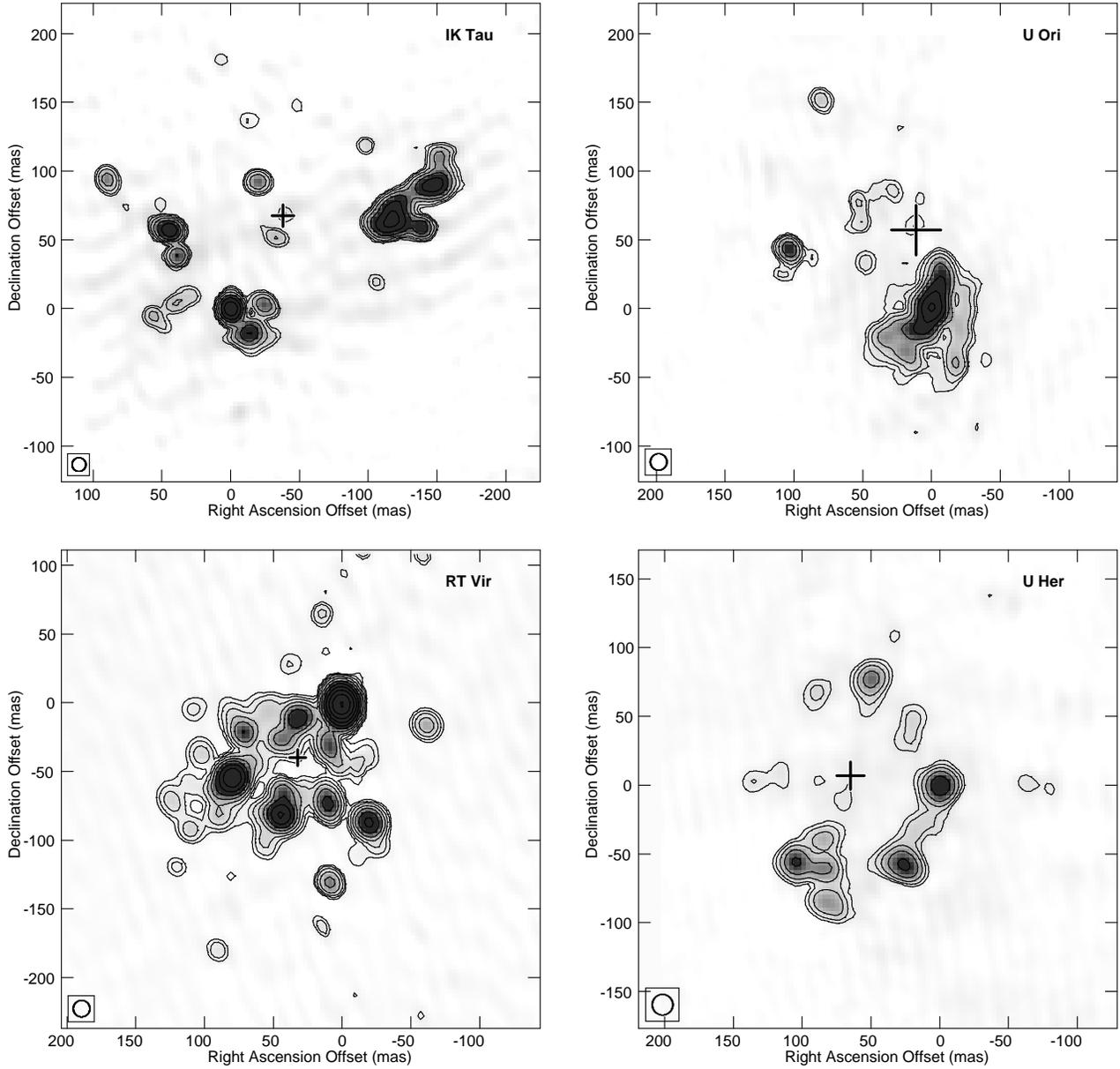

\resizebox{8.6cm}{!}{\epsfbox{IKTAU.SQAGPS}}
\resizebox{8.6cm}{!}{\epsfbox{UORI.SQAGPS}}
\resizebox{8.6cm}{!}{\epsfbox{RTVIR.SQAGPS}}
\resizebox{8.6cm}{!}{\epsfbox{UHER.SQAGPS}}
\caption{Integrated H$_{2}$O maser emission from four AGB stars 
observed by MERLIN in April 1994.  The origin
is at the position of the reference feature.  
The cross shows the estimated stellar position (Section~\ref{spos}).
The restoring beam is
shown in the lower left corner. The grey-scale shows all emission, the
contour plots were made after blanking each channel at its
$5\sigma_{\rm rms}$ level. The contour levels are at (1, 2, 4...)$
\times$ the lowest contour level. This is at 0.1, 0.05, 0.3 and 0.3 Jy
km s$^{-1}$ beam$^{-1}$ for IK Tau, U Ori, RT Vir and U Her 
respectively.}
\label{SQAGPS}
\end{figure*}

\begin{figure*}
\vspace*{-7.5cm}
\resizebox{17.1cm}{!}
{\rotatebox{-90}
{\epsfbox{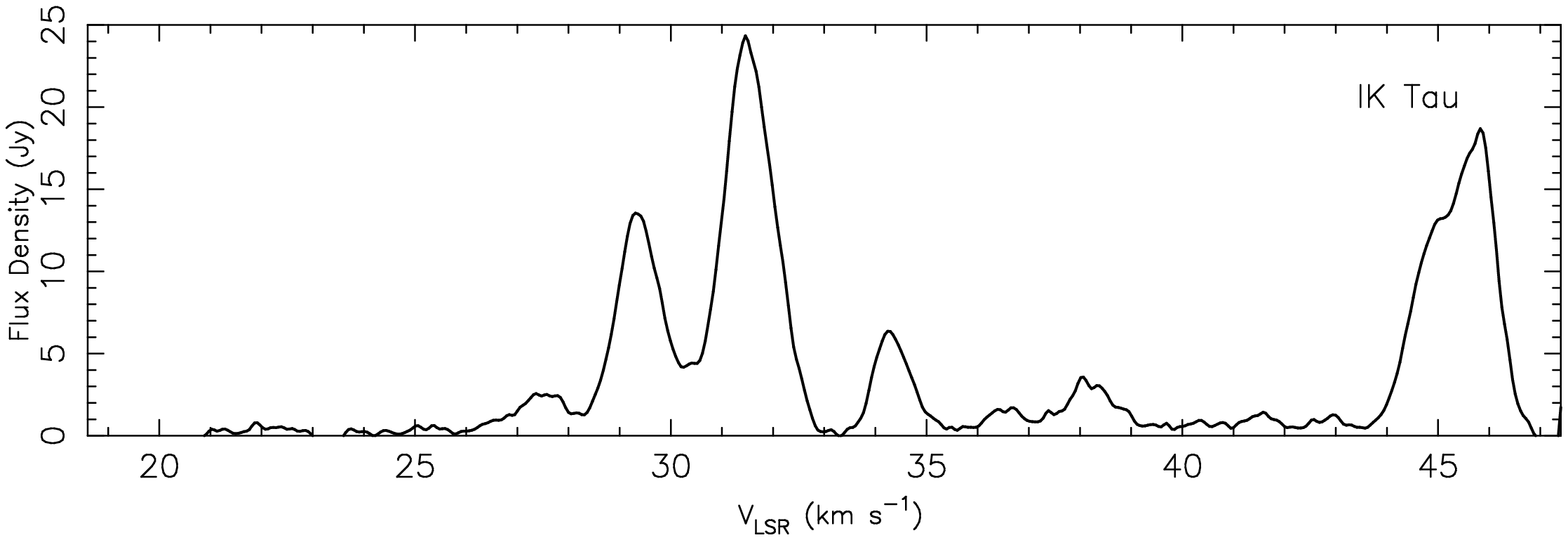}}
}
\hspace*{-1cm}
\vspace*{-0.6cm}
\resizebox{15cm}{!}{\epsfbox{IKTAU.KPS}}
\caption{{\em Top} MERLIN spectrum of IK Tau 22-GHz emission. 
{\em Bottom} Maps of emission integrated every 12 channels.  
The origin is at the position of the reference feature.  Each pane
is  labelled with the central velocity.  The contour levels are (1, 2,
4...) $\times$ 0.04 Jy bm$^{-1}$. The restoring beam is shown in the first pane.
 The panes containing the brightest emission (at
46.4, 45.1 and 31.2 km s$^{-1}$) contain some artefacts due to dynamic
range limitations.
}
\label{IKTAU}
\end{figure*}

\begin{figure*}
\vspace*{-7.5cm}
\resizebox{17.1cm}{!}
{\rotatebox{-90}
{\epsfbox{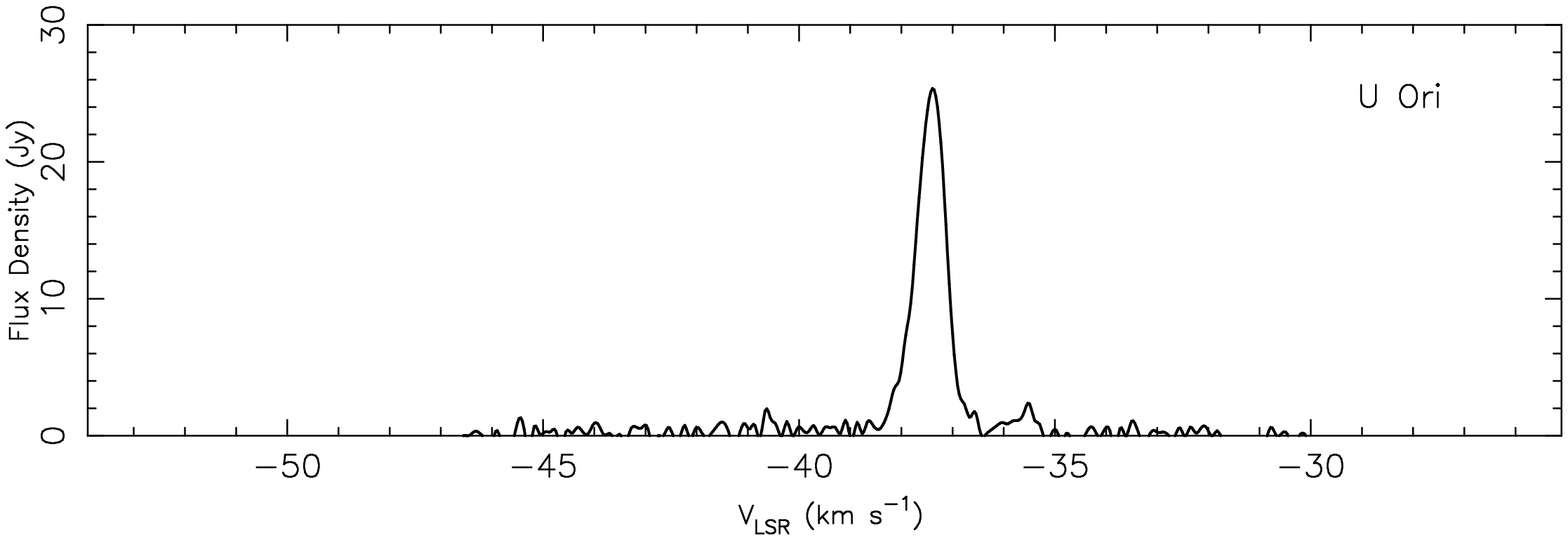}}
}
\hspace*{-1cm}
\vspace*{-0.6cm}
\resizebox{15cm}{!}{\epsfbox{UORI.KPS}}
\caption{{\em Top} MERLIN spectrum of U Ori 22-GHz emission. 
{\em Bottom} Maps of emission integrated every 10 channels. 
The origin is at the position of the reference feature.  Each pane
is  labelled with the central velocity.  The contour levels are (1, 2,
4...) $\times$ 0.04 Jy bm$^{-1}$. The restoring beam is shown in the first pane. 
The pane containing the brightest emission (at
--37.4 km s$^{-1}$) contains some artefacts due to dynamic
range limitations.
}
\label{UORI}
\end{figure*}

\begin{figure*}
\vspace*{-7.5cm}
\resizebox{17.1cm}{!}
{\rotatebox{-90}
{\epsfbox{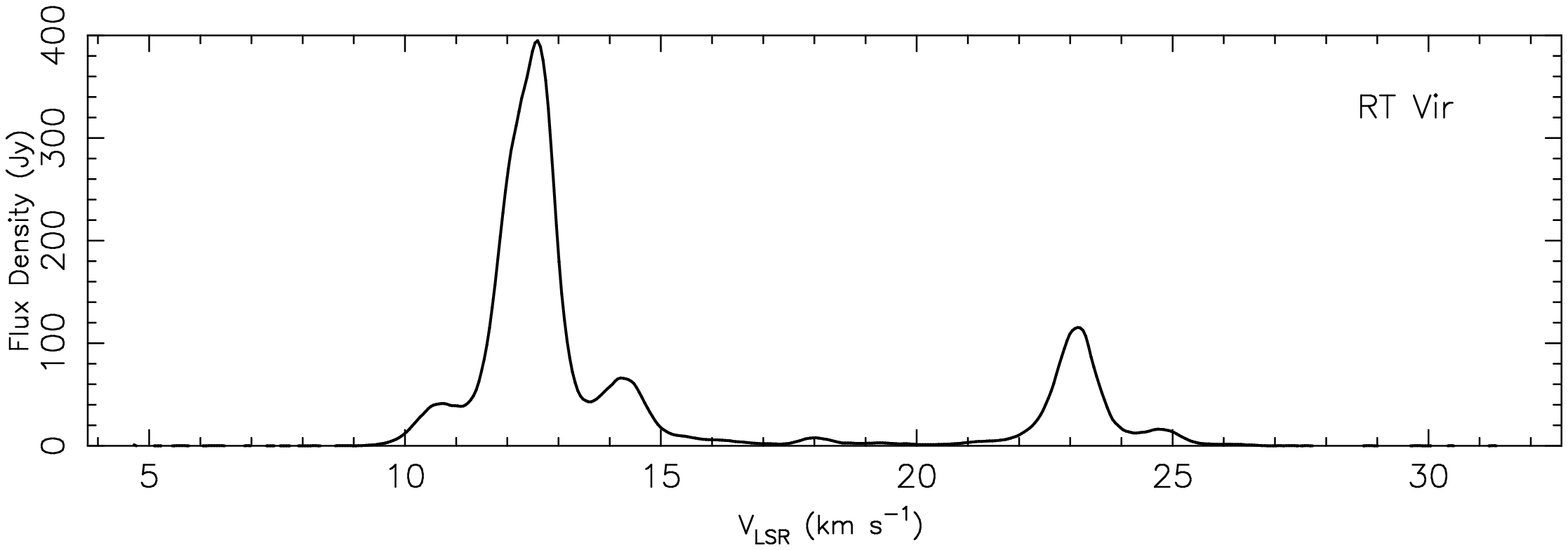}}
}
\hspace*{-1cm}
\vspace*{-0.6cm}
\resizebox{15cm}{!}{\epsfbox{RTVIR.KPS}}
\caption{{\em Top} MERLIN spectrum of RT Vir 22-GHz emission. 
{\em Bottom} Maps of emission integrated every 12 channels.  
The origin is at the position of the reference feature.  Each pane
is  labelled with the central velocity.  The contour levels are (1, 2,
4...) $\times$ 0.07 Jy bm$^{-1}$. The restoring beam is shown in the first pane.
The panes containing the brightest emission (at
23.2, 13.1 and 11.8 km s$^{-1}$) contain some artefacts due to dynamic
range limitations.
}
\label{RTVIR}
\end{figure*}

\begin{figure*}
\vspace*{-7.5cm}
\resizebox{17.1cm}{!}
{\rotatebox{-90}
{\epsfbox{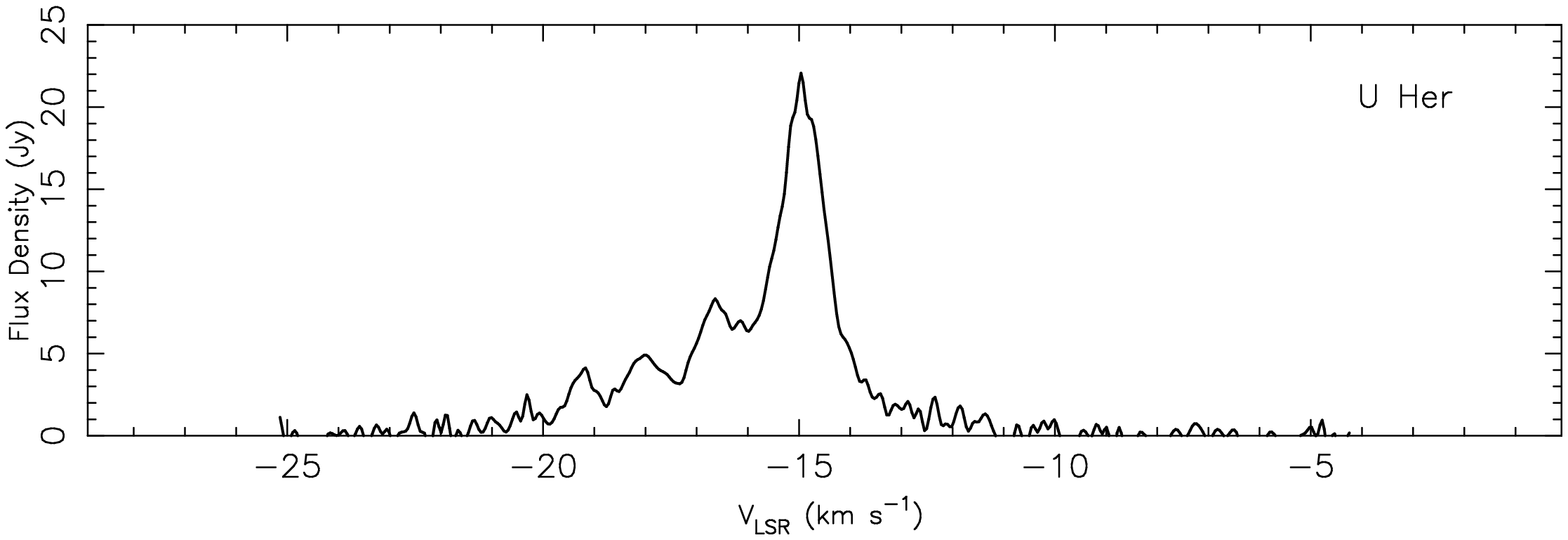}}
}
\hspace*{-1cm}
\vspace*{-0.6cm}
\resizebox{15cm}{!}{\epsfbox{UHER.KPS}}
\caption{{\em Top} MERLIN spectrum of U Her 22-GHz emission. 
{\em Bottom} Maps of emission integrated every 10 channels.  
The origin is at the position of the reference feature.  Each pane
is  labelled with the central velocity.  The contour levels are (1, 2,
4...) $\times$ 0.07 Jy bm$^{-1}$. The restoring beam is shown in the first pane.  
The panes containing the brightest emission (at
--14.6, --15.6 and --16.7 km s$^{-1}$) contain some artefacts, mostly
near the centre, due to dynamic
range limitations.
}
\label{UHER}
\end{figure*}

\begin{figure*}
\hspace*{-0.3cm}
\rotatebox{-90}{\resizebox{19.5cm}{!}{\epsfbox{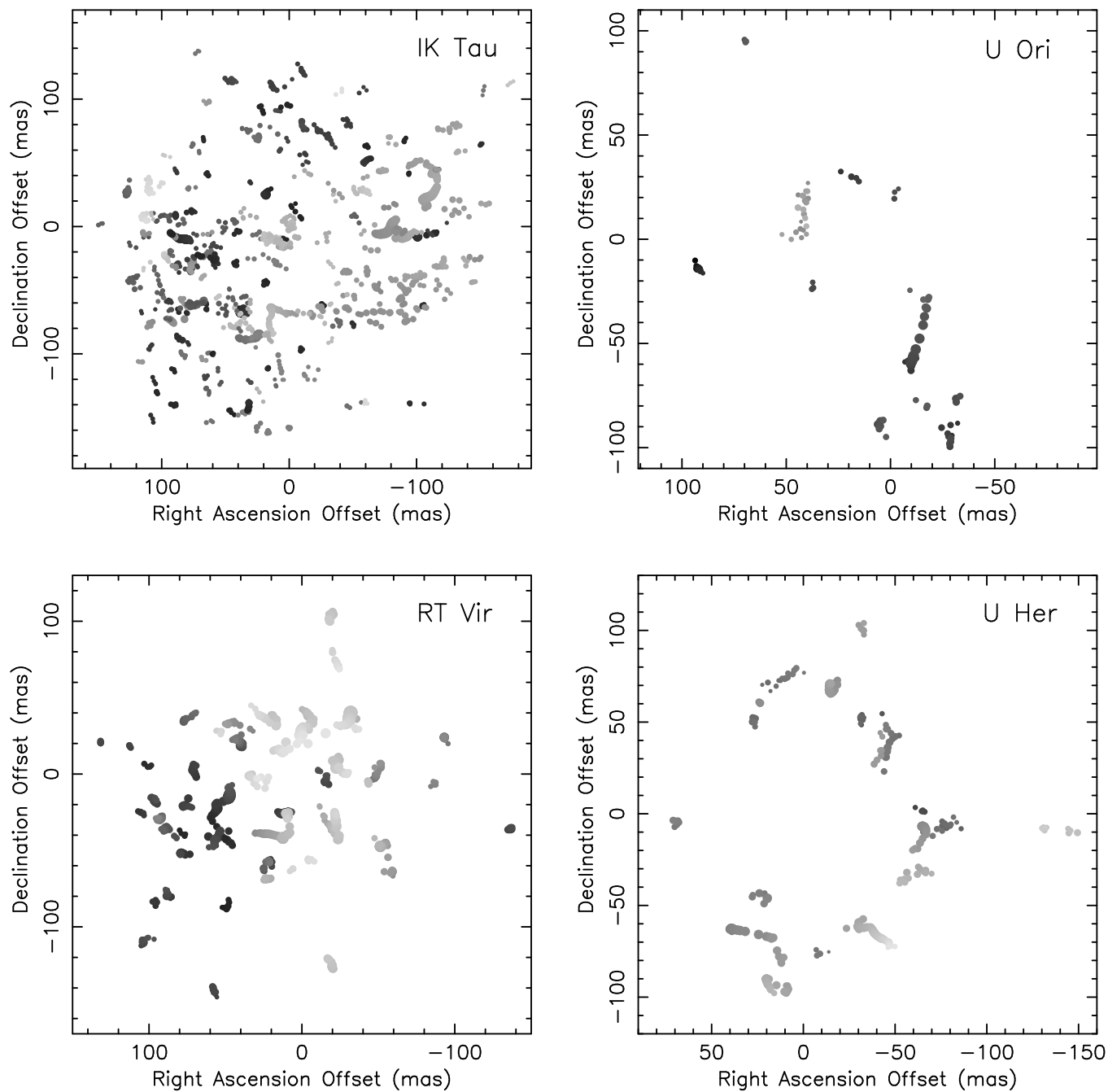}}}
\caption{The positions of the H$_{2}$O maser components for each
source.  Each panel is labelled with the source name. The origin is at
the estimated stellar position.  The diameter of each spot is
proportional to the logarithm of its flux density.  The paler spots
correspond to more blue-shifted emission and the darker spots
correspond to more red-shifted emission.}
\label{ALL.GPS}
\end{figure*}

\subsection{MERLIN spectra and maps}
\label{maps}

Maps of the integrated 22-GHz emission from each source are shown in
Figure~\ref{SQAGPS}.  The emission comes in each case from elliptical
regions $\sim$0.3~arcsec in extent, in which are embedded hot spots
$\sim$20~mas in extent which contribute most of the flux.
Table~\ref{peaks} lists for each source the total angular extent of
the emission detected, at the 5$\sigma_{\rm rms}$ level
$\Delta\theta_{\rm tot}$ (column(2)), together with the stellar
velocity (column (3)), the minimum and maximum of Doppler velocity
(columns (4) -- (5)) and the range $\Delta V_{\rm LSR}$ (column (6)),
the peak flux density $S_{\rm peak}$ (column (7)) and the integrated
emission $S_{\rm tot}$ (column (8)).  Figure~\ref{SQAGPS} also shows
estimated stellar positions derived in Section~\ref{spos}.  It is clear
that in each case the brightest emission is mostly found towards the
outer parts of the H$_{2}$O envelope, with little emission from the
direction of the star.

Figs.~\ref{IKTAU} to~\ref{UHER} show the results in 
more detail:  a MERLIN spectrum for each source, 
together with maps of the emission integrated over velocity ranges of 
$\sim$1.2~km~s$^{-1}$.  
Individual channel maps contain mostly well-separated, slightly resolved
patches of emission.
Comparison with single dish data (autocorrelations and
independent monitoring e.g. \pcite{Rudnitskij00}) shows that MERLIN
detected all the flux to within the errors ($\le10$ percent).  
Note however that the extreme red-shifted emission from IK Tau
extends right up to (and possibly beyond) the edge of our observing
band. 
Henceforth red- and blue-shifted are used to denote emission with
$V_{\rm LSR} > V_*$ and $V_{\rm LSR} < V_*$ respectively.  

Fig.~\ref{ALL.GPS} shows the positions of the fitted maser components
in each channel, with grey-scale used to 
indicate velocity.  In Section~\ref{analysis} we analyse 
the properties of the individual maser components.  
Here we concentrate on the global aspects of the 
maser distributions around each star. 
Results for each source are given in the following four Sections.

\subsection{IK Tau}
\label{IK}
IK Tau has the most complex spectrum and the widest velocity range 
of the four sources we studied.  There are two bright spectral peaks 
asymmetrically distributed on either side of the stellar velocity:  
the blue-shifted peak by 6~km~s$^{-1}$ 
and the red-shifted peak by 12~km~s$^{-1}$. 
The channel maps of IK~Tau show that these bright emission peaks 
are due to multiple bright  
clumps spread over an elliptical region $\sim200$~mas E-W 
and $\sim100$~mas N-S.  Fainter
clumps of emission are found in all channels, from 
an almost spherical region.  Moderately blue-shifted emission to
the west tends to be brighter than that to the east; the opposite is true
for red-shifted emission.  At velocities close to $V_*$ a clump to the
south dominates the emission.  The extreme red- and blue-shifted emission
features are offset towards the east.  

Previous MERLIN observations by \scite{Yates94} in 1985 
detected only the 
bright elliptical region of emission but not the fainter spherical 
envelope.  
No individual maser features from the earlier MERLIN maps 
can be matched to features in the new maps.  
On the other hand, 
\scite{Yates94} found a close correspondence between MERLIN and VLA 
maps made 16 months apart (\pcite{Lane87}; \pcite{Bowers93}).  
Many maser features had survived over that time period and  
\scite{Yates94} were able to measure expansion of the H$_2$O maser envelope.  

Although the overall appearance of the envelope in 1994 is of a
spherically symmetric shell, nevertheless there is a clear offset 
between the near and far sides of the shell, with red-shifted emission 
mainly to the  east and blue-shifted emission mainly to the west.  
This is particularly noticeable in Figure~\ref{ALL.GPS}.  
The E-W velocity segregation can be explained if the shell has an 
equatorial density enhancement so that the brightest masers lie in an
oblate spheroid.  The plane of the equator is at an angle of
inclination $i$ to the line of sight, with the eastern end of the polar axis
approaching us.  \scite{Hale97} report that 3.1~$\mu$m speckle
interferometry shows an approximately circular disc of warm dust 140
mas in diameter with a resolution of 40 mas.  If this is the
projection of the inner regions of the H$_2$O maser shell, with a
minimum axial ratio within the uncertainty, this constrains
$i\ga 45\degr$.
\scite{Bowers91} fig.~2 shows model maps for
 an oblate spheroid at $i=45\degr$ producing an asymmetric
appearance at velocities around midway between $V_*$ and the 
extreme velocities.  
This model does not include acceleration, which produces brighter
tangential beaming.
 \scite{Kirrane87} found that the OH mainline masers trace a biconical
outflow with a wide opening angle, surrounded by OH-1612 MHz masers in
the equatorial region, with $i\approx60\degr$.

\subsection{U Ori}
\label{UO}
U Ori has the narrowest velocity range of the four sources, 
with a single dominant spectral peak at a red-shift of 2~km~s$^{-1}$
and mainly red-shifted emission.  
The brightest emission, at
$-37.4$~km~s$^{-1}$, comes from an extended region to the south west, 
while the remainder of the emission is to the north and east.  
The distribution as a whole is elongated NE-SW, at a position 
angle of $\sim$30\degr.
The maser envelope is the most sparsely filled of the four we 
studied (Figure~\ref{ALL.GPS}).

U Ori has been monitored at 22 GHz by \scite{Rudnitskij00} since 1980
using the Pushchino single dish radio telescope.  
They observed H$_{2}$O maser flares, which they propose 
arise due to shocks propagating radially out from the star.
Spectacular OH maser flares have also been observed (\pcite{Pataki74}; \pcite{Chapman85} and references therein).

\scite{Bowers94} mapped U Ori in 1988 using the VLA and found a 
similar elongated distribution at position angle $\sim$60\degr.  
The major difference between our maps and theirs is that they 
detected weak blue-shifted emission out to --49~~km~s$^{-1}$ 
near the centre of the distribution, close to our estimated 
stellar position. 
The position angle of the H$_{2}$O envelope 
observed by \scite{Bowers94} in 1988 is identical to 
the position angle of the OH 1612-MHz maser flare 
measured by Chapman \& Cohen (1985) in 1983.

\subsection{RT Vir}
\label{RT}
RT Vir has the greatest 22-GHz peak flux density and total flux 
of the four sources, with two bright spectral peaks 
$\sim$5~km~s$^{-1}$ each side of the stellar velocity.  

The emission has an angular extent $\le370$ mas,
containing an inner ring of bright masers $\sim150$
mas in diameter.  As in IK~Tau the emission contributing to the 
strongest spectral peaks is due to many compact maser regions 
spread over a large region.  The total extent of the emission 
region is greatest near the stellar velocity and least near the 
extreme velocities, as expected for an expanding spherical shell, 
but again as in IK~Tau the extreme red- and blue-shifted features 
are displaced from each other.  
Extreme blue-shifted emission from RT Vir comes from near the centre
of the channel maps, but the extreme red-shifted emission is to 
the south east.  
The most extended emission consists of
faint patches to the N (slightly blue-shifted) and to the
south (slightly red- and blue-shifted).

The integrated emission shown in Figure~\ref{SQAGPS} gives the 
impression of a well-filled spherically symmetrical shell, but in 
fact there is a systematic displacement between the near and far 
sides of the envelope, with most  red-shifted masers to the east and 
blue-shifted masers to the west (Figure~\ref{ALL.GPS}  
This E-W velocity segregation can be explained if the shell has an 
equatorial density enhancement so that the brightest masers lie in an
oblate spheroid.  

Previous maps of RT~Vir show a similar displacement between the 
near and far sides of the H$_{2}$O envelope (Imai et al. 1997, 
\pcite{Bowers94},   \pcite{Yates94}, 
Bowers, Claussen \& Johnston 1993).
Despite this similarity, individual maser features cannot be matched.  

RT Vir has been monitored at 22 GHz at Pushchino since 1985
(\pcite{Mendoza-Torres97}; \pcite{Lekht99}). It is extremely rapidly
variable, with individual flares lasting less than 3 months.   
The MERLIN spectrum has a peak 150 Jy brighter than measured at Pushchino 
a month earlier. 

RT Vir appears to have an equatorial density 
enhancement in its 22-GHz maser distribution such that the brightest
emission comes from an oblate spheroid tilted in the plane of the sky
so that the western side is approaching us.  This structure
 persists much
longer than the lifetime of identifiable individual maser clouds. 
The model maps of \scite{Bowers91} fig.~2 qualitatively illustrate the
appearance of the brightest emission (as for IK Tau). 
 The polar axis is likely to be at a
projected angle similar to the direction of the red-blue velocity
offset.

\subsection{U Her}
\label{UH}
The emission from U~Her is strongest near the stellar velocity and 
mainly blue-shifted.  The emission comes from an elliptical  
ring-like region centred on our estimated stellar position, and 
elongated roughly north-south.   
The brightest emission lies to the south and west.  
The emission from each channel map is likewise ring-like, with 
no systematic change of radius with velocity.  The most complete 
rings are seen near the stellar velocity, whereas at the extreme 
velocities just a single emission region is seen, offset from the 
stellar position by $\sim$80~mas in each case.  
The most red-shifted emission occurs near the inner edge of the ring
towards the west.  The extreme blue-shifted emission lies to the 
south south west.  The only emission clearly outside the ring is a 
moderately blue-shifted spur $\sim100$ mas to the west.

A similar ring-like distribution was observed by  \scite{Bowers94} in 1988
with the VLA, when the source was an order of magnitude brighter. 
However there is no detailed correspondence of 
individual maser features between the two epochs.  
The extreme blue-shifted emission which formed 
the north-eastern part of the ring in 1988 has now faded.  
The radius of the ring may appear to have increased by $\sim$20~mas 
in the 5 years between the two set of observations, 
but this result is only marginally significant given the 
70-mas beam of the VLA and the lack of detailed agreement at the 
level of individual components.  
\scite{Colomer00} also found a ring distribution using the 
VLA in 1990, but again with no detailed correspondence between 
individual maser features at the different epochs.  

Previous MERLIN measurements by \scite{Yates94} 
in 1985 detected only a single extended region of emission near 
the stellar velocity.  

The strong differences in appearance at different epochs show that
individual clouds in the envelope of U~Her cannot
  be followed for more than $\sim$2 stellar periods.  This is consistent with
the shock model discussed later in Section~\ref{amp}.  U~Her, like
U~Ori, has also undergone OH maser flares \cite{Etoka97}, which may be
another result of shocks.

\section{The dynamics of the circumstellar envelopes}
\label{discuss}

\subsection{Stellar position}
\label{spos}

Much of our further analysis requires an estimate of the stellar position.  
We used two methods to estimate the centre of expansion ($X$, $Y$), 
assumed to be the stellar position.  Both methods assume that $V_*$ is 
accurately known and use least-squares minimisation routines.  

The shell-fitting method of \scite{Yates93} finds the point which is
most nearly equidistant from all components in angular separation and
in velocity.  This assumes the maser components have a point-symmetric
distribution about the centre of a spherical shell.  However for
non-spherical shells 
it is more reliable to use a method based on the
quenching radius (Section~\ref{quench}). Values of ($X$, $Y$) are
found which maximise the angular separation from all components with
velocities within the range $V_*\pm(\Delta V_{\rm LSR}/n)$,  
where $n$ is between 4 and 8 (see Table~\ref{peaks}).

In practice, similar results were obtained using both methods for IK
Tau, RT Vir and U Her.  For U Ori the shell-fitting method did not
converge, but the quenching radius method gave consistent results
for  different values of $n$.  Table~\ref{starpos} gives ($X$, $Y$)
found by the quenching radius method.
 The error quoted is the
difference between this position and the mean position found by all
methods.

Figs.~\ref{IKTAU} to ~\ref{UHER} ({\em bottom}) show that in all four
CSEs either or both of the extreme blue- and red-shifted features are
offset from the assumed stellar position by 20 -- 115 mas.  We did not
consider the mid-point of the extreme blue- and red-shifted features
as the stellar position as in every case this would be very asymmetric
with respect to most of the masers.  The offsets could be explained by
turbulence of 1.1 -- 1.6 km~s$^{-1}$ (\pcite{Chapman94};
\pcite{Chapman85t}), by clumpiness, or by more systematic causes
discussed in Sections~\ref{maps} and~\ref{amp}.

\begin{table}
\caption{Estimated stellar positions ($X$, $Y$) with respect to the reference feature 
for each source.}
\label{starpos}
\begin{tabular}{lll}
\hline
Source &($X$, $Y$)& Error in $X$ and $Y$\\
	&(mas)	&(mas)	\\
\hline
IK Tau & ($-38$, $67$) & 4\\
U Ori	&($11$, $57$) &9\\
RT Vir&($32$, $-40$) &3\\
U Her& ($65$, $7$) & 5\\
\hline
\end{tabular}
\end{table}

\subsection{Kinematics of the H$_{2}$O  shells}
\label{limits}

\begin{figure*}
\hspace*{-0.3cm}
\rotatebox{-90}{\resizebox{19.5cm}{!}{\epsfbox{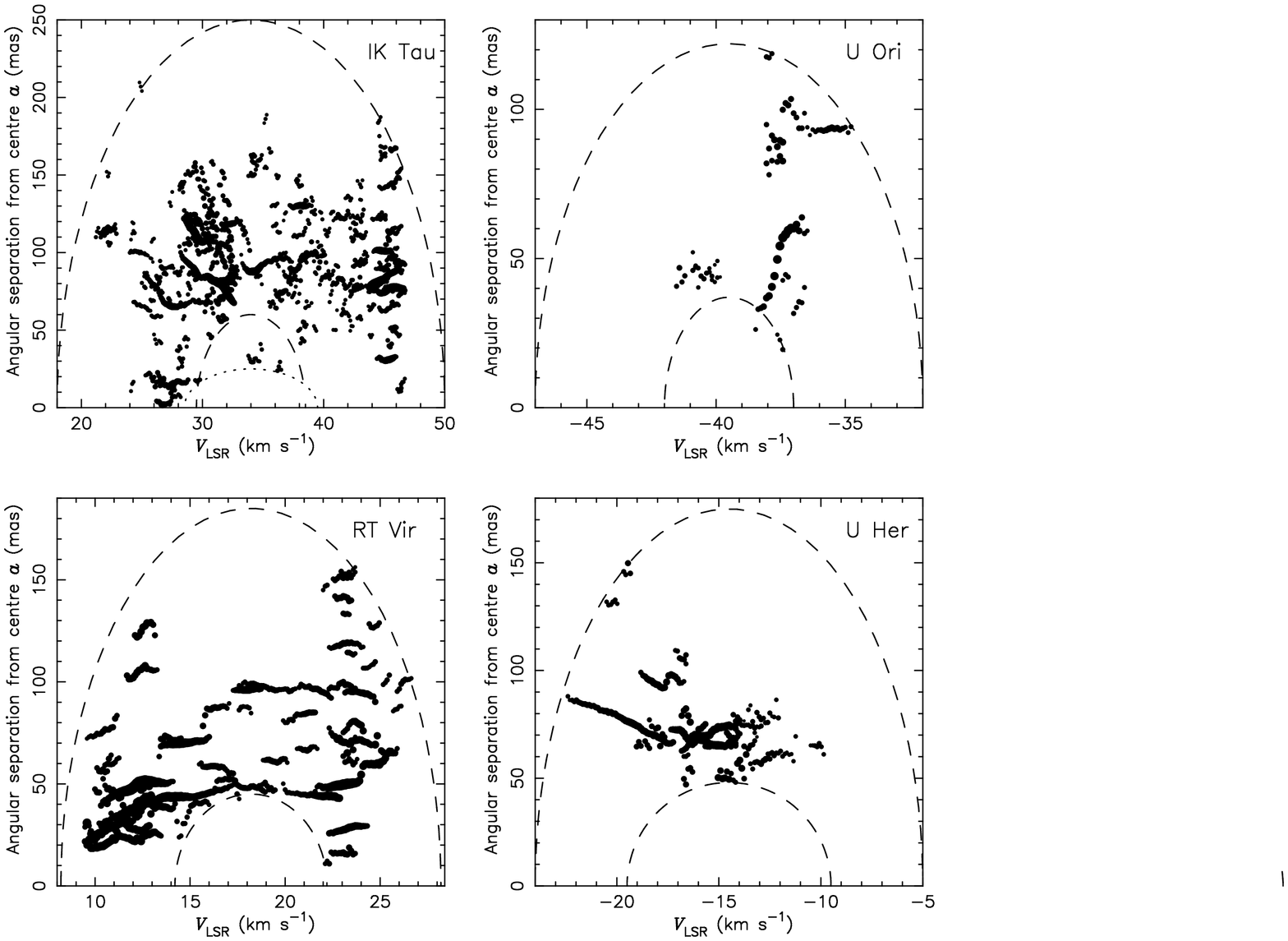}}}
\caption{Radius-velocity plots of the 22-GHz masers around IK Tau, 
U Ori, RT Vir and U Her, showing the   
angular separation $a$ from the assumed  stellar position, 
as a function of $V_{\rm LSR}$. 
The diameter of each spot
shown proportional to the logarithm of its flux density. The dashed lines
show ellipses centred on $V_*$ which were fitted by eye to the inner
and outer limits of the maser distribution.   The dotted line shows
an alternative inner limit for IK Tau (see Section~\protect{\ref{quench}}).}
\label{RV}
\end{figure*}

Previous MERLIN observations of circumstellar H$_{2}$O masers 
have suggested a thick-shell model, in which the masers are distributed 
irregularly but the velocity field is regular and spherically 
symmetric, increasing steadily with radius through the 
H$_{2}$O maser region 
(\pcite{Yates94}, \pcite{Richards99} and references therein).  
That model also describes the present observations well.  In 
Fig.~\ref{RV} we show the radius-velocity plots for the  
four sources.  The inner and outer 
edges of the H$_{2}$O envelope were estimated by fitting the dashed 
ellipses by eye to the data.  From these we obtain the inner 
and outer radii of the H$_{2}$O maser zone,  
which we denote by $r_{\rm i}$ and $r_{\rm o}$, together with 
the corresponding expansion velocities $v_{\rm i}$ and $v_{\rm o}$.   
These parameters are listed in columns (2) -- (7) of
Table~\ref{rvtab}.  
The error in the estimated stellar position (Section~\ref{spos}) 
is applicable to the estimates of $r_{\rm i}$ and $r_{\rm o}$ and 
is given in column (8) of the Table.  

Two alternative inner limits are shown for IK Tau, since the limit (a)
with the larger value of $r_{\rm i}$ is well defined but a few
features lie within this giving a smaller value (b). The limits for U
Ori and U Her are more uncertain but have been informed by the
appearance of these stars at other epochs (e.g. \scite{Bowers94},
Murakawa priv. comm.) when emission appeared at different position
angles and/or velocities.  

In all cases the expansion velocity is less than the escape 
velocity at the inner edge of the H$_{2}$O maser zone, but 
easily exceeds the escape velocity at the outer edge of the 
H$_{2}$O maser envelope.  This was already established by 
\scite{Yates94} for IK Tau and RT Vir, and seems to be a 
characteristic of circumstellar  H$_{2}$O envelopes in general.

The shell limits were used to derive the logarithmic velocity gradient
$\epsilon=d(\ln v)/d(\ln r)$ in each circumstellar envelope.  
These are listed in column (9) of Table~\ref{rvtab}, along with the error 
 $\sigma_{\epsilon}$, the linear 
velocity gradient $K_{\rm v}$ and its uncertainty $\sigma_{\rm K_{\rm v}}$
 (in columns (10) to (12) respectively).  
These are average values for the H$_{2}$O maser shells as there are
local variations in each CSE. 
Acceleration is strongest nearer the star in each case.  

The values of  $\epsilon$ are similar in the four stars:  they range from 
0.5 to 0.9, and are very similar to the values derived from MERLIN 
observations of supergiants (\pcite{Yates94} and references therein).  
These are the first direct measurements of $\epsilon$ for low-mass 
AGB stars, since ours are  the first measurements to resolve the 
thickness of the H$_{2}$O shells.  Previous MERLIN measurements by 
\scite{Yates94} made at lower angular resolution (without the 
Cambridge telescope) could only set 
upper limits on  $\epsilon$ for IK Tau and RT Vir.  

The total H$_{2}$O maser luminosity $\Phi$  (equivalent isotropic
luminosity)  is given in column (13) of 
Table~\ref{rvtab} for completeness.  Although the three Miras
all have peaks of similar intensity (Table~\ref{peaks}) and are at a
similar distance, U Ori has only about a quarter of the luminosity of
IK Tau, U Her being at an intermediate value.  RT Vir is almost twice as
luminous as the brightest of the Miras.

The present measurements are sufficiently detailed to allow us to
investigate the variation of maser emissivity with radius in the CSEs.
Figs.~\ref{RV} and the spectra in Figs~\ref{IKTAU} -- {UHER} show that
more and brighter masers were detected in the inner parts of the
shells.  This is broadly consistent with the models of \scite{Cooke85}
who predict a sharp increase in maser brightness near $r_{\rm i}$ and
a gradual decline towards $r_{\rm o}$.

\begin{figure}
\hspace*{-0.2cm}
\rotatebox{270}{
\resizebox{9cm}{!}{\epsfbox{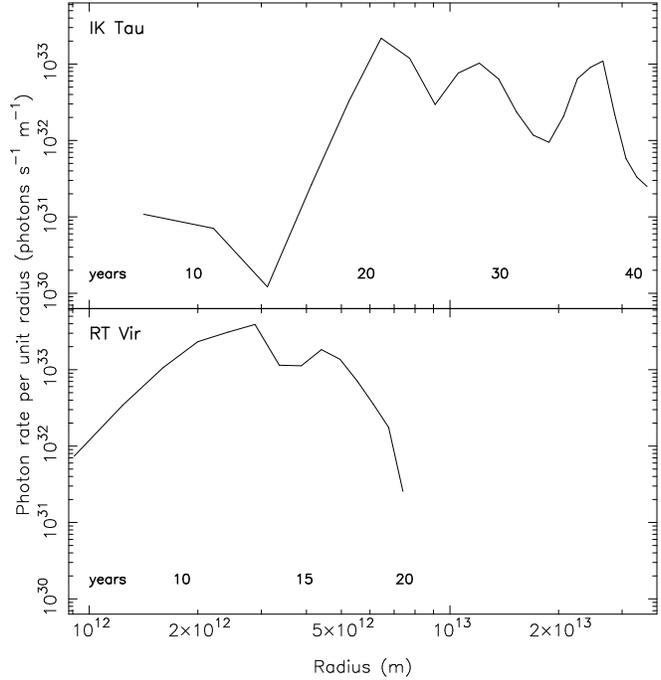}}}
\caption{The 22-GHz photon rate per unit radius for IK~Tau and RT~Vir,
as a function of radial distance from the assumed stellar position.
The bold numbers above the abscissa are estimates of the time taken
for material in the stellar wind to reach that distance, see Section~\protect{\ref{quench}}.}
\label{DPHIDR.PS}
\end{figure}

The 22-GHz maser shells around IK~Tau and RT~Vir are well-filled
enough for us to estimate the photon luminosity as a function of distance
from the star.  We used the model of a spherically symmetric velocity
field within the shell limits given in Table~\ref{rvtab} to assign
each maser component a full set of vectors to describe its position
and velocity in all 3 spatial dimensions, as described in Murakawa et
al. (2002, MNRAS, submitted). We then calculated the flux density in
sub-shells with a cross-section in the plane of the sky of 10 mas, and
used this to find the 22-GHz photon rate per unit radius $d\Phi/d{r}$.
This is shown in Fig.~\ref{DPHIDR.PS}.  The results are robust at 
the 10 percent level to departures from spherical symmetry by up to a 
factor 2 in  $\epsilon$ \cite{Richards99}.

The results shown in Fig.~\ref{DPHIDR.PS} were compared with the model
results of \scite{Cooke85}.  Initially we used the mass loss rates
$\dot{M}_{\rm CO}$ given in Table~\ref{sources}.  For IK Tau
$d\Phi/d{r}$ reaches a maximum value of $\sim2\times10^{33}$ photons
s$^{-1}$ m$^{-1}$ at a radius of $6\times10^{12}$ m. This distance
is within the range predicted by the models of \scite{Cooke85} for the
given $\dot{M}_{\rm CO}$.  However, for RT~Vir, the maximum
$d\Phi/d{r}\sim4\times10^{33}$ photons s$^{-1}$ m$^{-1}$
 occurs at a radius of
$3\times10^{12}$ m, which is $\sim3$ times the radius predicted by
\scite{Cooke85} for the given $\dot{M}_{\rm CO}$.

We return to this comparison later in Section~\ref{amp}, 
when we discuss the over-density of the  H$_{2}$O maser clouds 
with respect to their surroundings.

\begin{table*}
\caption{Parameters of the H$_{2}$O maser envelopes (see
Sections~{\protect{\ref{limits}}} and {\protect{\ref{shells}}} for details).
}
\begin{tabular}{lrrrrrrrrrrrc}
\hline
Source	&\mc{2}{c}{$r_{\rm i}$} & $v_{\rm i}$ &\mc{2}{c}{$r_{\rm o}$}& $v_{\rm o}$&$\sigma{\rm r}$ &$\epsilon$&$\sigma_{\epsilon}$&$K_{\rm v}$&$\sigma_{\rm K_{\rm v}}$&$\Phi$ \\
&(mas)&(au)	& (km s$^{-1}$)&(mas)    &(au)	& (km s$^{-1}$)&(au)  &&&\mc{2}{l}{($10^{-12}$} &\mc{1}{l}{($10^{42}$}\\
&&&&&&&&&&\mc{2}{r}{km s$^{-1}$ m$^{-1}$)} &photons s$^{-1}$)\\
(1)&(2)&(3)&(4)&(5)&(6)&(7)&(8) &(9)&(10) &(11)&(12) &(13)\\ 
\hline
IK Tau(a) &60 	&16.0 	&4.5 &250	&66.5	& 16.0	& 1.0	& 0.89	&0.08	&1.52&0.06&4.63	\\
IK Tau(b) &25	&6.7	&5.5 &250	&66.5	& 16.0	& 1.0	& 0.46	&0.09	&1.17&0.04&4.63	\\
U Ori 	  &37	&9.8	& 2.5&120 	&32.5 	&7.5 	&2.4 	&0.92	& 0.24	&1.48&0.23&0.85	\\
RT Vir 	  &45	&6.0 	&4.0 &185	& 25.6 	& 10.0	& 0.4 	&0.65 	&0.05 	&2.15&0.08&7.88	\\
U Her 	  &48	&12.8   &5.0 &175	&46.6  	&9.5	&1.1	& 0.50 	& 0.04	&0.89&0.05&1.92	\\
\hline
\end{tabular}
\label{rvtab}
\end{table*}

\subsection{The velocity fields of the stellar winds}
\label{shells}

\begin{figure}
\hspace*{-0.3cm}
\rotatebox{-90}{\resizebox{13cm}{!}{\epsfbox{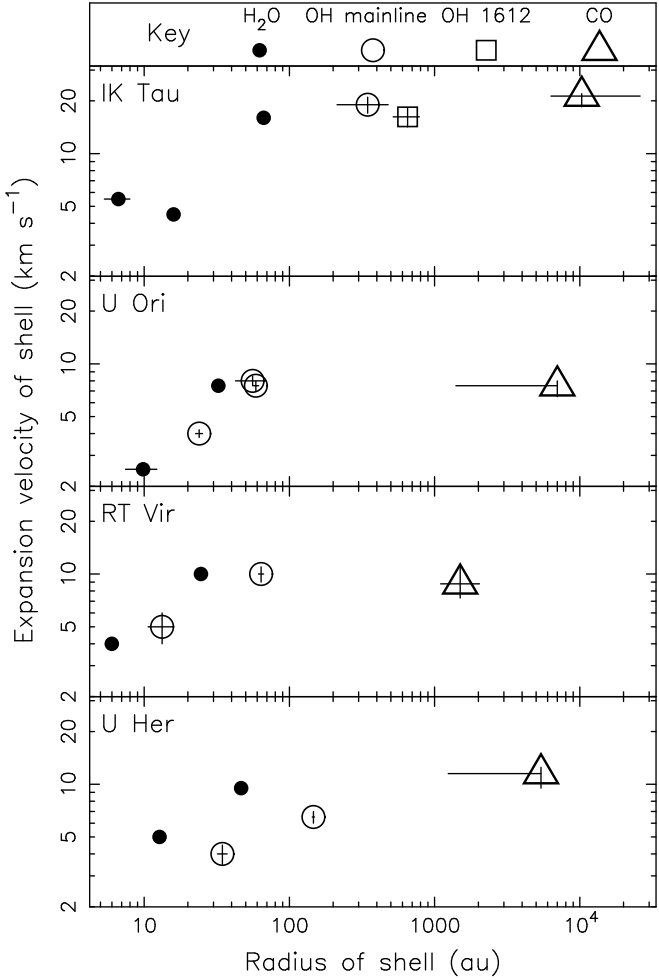}}}
\caption{Expansion velocity as a function of radial distance 
from the four stars IK Tau, U Ori, RT Vir and U Her.  The H$_{2}$O
data are from Table~\protect{\ref{rvtab}}.
The OH data are from references given in 
Section~\protect{\ref{shells}}.
The CO values are outer limits from references
 given in Table~\ref{sources}.}
\label{ALLRVE}
\end{figure}

Fig.~\ref{ALLRVE} shows plots of expansion velocity vs. radial
distance for the four stars, compiled using our H$_{2}$O data together
with data on other molecular species taken from the literature.  All
four stars have bright OH mainline masers which have been mapped
(\pcite{Chapman91}; \pcite{Chapman94}; \pcite{Bowers88};
\pcite{Richards00o}; \pcite{Kirrane87}), while IK Tau was mapped at
1612~MHz by \scite{Kirrane87}.  The OH 1612-MHz maser flares in U Ori
were imaged by \scite{Chapman85} but as
there was no clear shell structure there was no obvious way to assign a radial
distance to the emission.  CO data were taken from the references in
Table~\ref{sources}.  Only RT Vir is well resolved in CO
\cite{Stanek95}.  The values given for the Miras are based on
different models used by the various authors, but they also observed
some of the same stars as \scite{Stanek95}.  Comparison of the results
for objects in common suggests a CO diameter of 10000 au for IK Tau,
and sizes a factor of $3 - 5 \times$ smaller than those given for U
Her and U Ori.  The error bars for CO data in Fig.~\ref{ALLRVE}
reflect this.  The distances were taken as 133 pc for RT Vir and 266
pc for the other stars.

Fig.~\ref{ALLRVE} show several striking features:
\begin{itemize} 
\item{There is a general increase of expansion velocity with
distance from the star, with some scatter. 
The strongest acceleration is in the region out to 30 au in the 
smallest CSE, RT Vir, and 70 au in the largest, IK Tau.}  
\item{IK Tau and U Her also show evidence for gentler acceleration 
continuing further  
outwards to $\sim$500 au, and perhaps as far as the CO region at 
$\sim$3000 au.}  
\item{The best filled 22-GHz maser shells, around IK Tau and RT Vir,
attain the greatest H$_{2}$O velocities and contain  faster
emission at a given radius than do the shells of U Ori and U Her.}
\item{In U Ori, U Her and RT Vir, the OH mainline masers 
 overlap the H$_2$O shells in angular
separations and velocities.  However the H$_2$O masers appear to have
greater velocities than OH masers at a similar radius.}
\end{itemize}

 \scite{Chapman86} observed acceleration of the wind from VX Sgr out
to at least 100 $R_*$ and showed that could be explained by an
increase in the dust absorption efficiency $\kappa_{\rm D}$ during the
outflow.  A similar kinematic pattern has been observed in other RSG
e.g. S Per \cite{Richards99} but hitherto it has been hard to confirm
similar behaviour in the smaller CSEs of Miras. 
Fig.~\ref{ALLRVE} and Table~\ref{rvtab} show that acceleration occurs
out to many tens of $R_*$ in these stars, and much further for two of
the stars: IK Tau and U Her. 

If dust properties are assumed to be constant \scite{Elitzur01} predict that
some acceleration does occur due to optical depth effects but terminal
velocity is attained at $\la10R_*$ and wind speed increases
monotonically with $\dot{M}$ for silicate dust \cite{Netzer93}. 
Monnier, Geballe \& Danchi (1998, 1999) \nocite{Monnier98}
\nocite{Monnier99} found periodic and long-term fluctuations in the
shape and intensity of the 9.7$\mu$m silicate feature towards IK Tau,
U Ori and U Her.  This suggests changes in the dust optical constants;
possible mechanisms include annealing or solid diffusion
(\scite{Gail99} and references therin), which could improve the
absorption efficiency at greater $r$.

Multi-epoch monitoring is continuing to investigate the effects of
stellar variability on the velocity field in CSEs.  \scite{Engels97}
found strong variability (``mode-switching'') in the 22-GHz line shape
and velocity width towards OH 39.7+1.5, with no evidence for velocity
gradients at large radial distances.  However, unlike the stars in our
sample, this object has cool IRAS colours and a 9.7-$\mu$m silicate
feature in absorption. \scite{Engels88} and \scite{Szymczak95} also
found that H$_{2}$O masers attain terminal velocity in objects with
high $\dot{M}$, whereas H$_2$O masers in thinner-shelled CSEs do not.
This is consistent with our results and suggests that stronger
acceleration is favoured when starlight can more easily penetrate the
CSE.

\section{The properties of maser clouds}
\label{analysis}

\begin{figure}
\hspace*{-0.7cm}
\rotatebox{270}{
\resizebox{5.5cm}{!}{\epsfbox{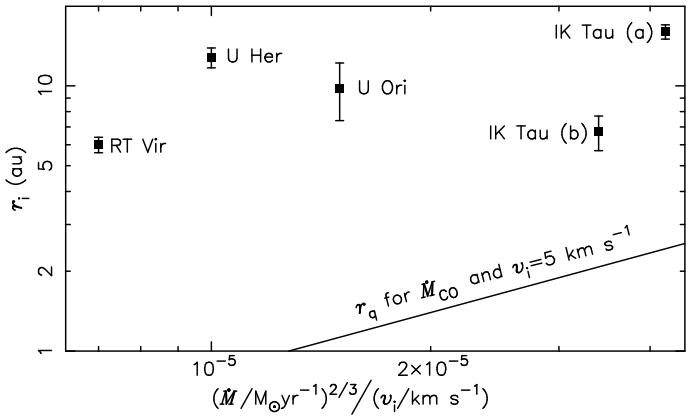}}}
\caption{The labelled points show the measured values of $r_{\rm i}$
as a function of $(\dot{M}_{\rm CO})^{2/3}/v_{\rm i}$.  The solid line
is the predicted $r_{\rm q}$ for a similar range of $\dot{M}_{\rm CO}$
at a typical $v_{\rm i}=5$~km~s$^{-1}$ and $f=4\times10^{-4}$.}
\label{MDOT}
\end{figure}

\subsection{Maser component sizes and temperatures}
\label{components}

We now analyse the properties of individual water vapour maser clouds.  
We begin by considering the Gaussian components fitted to each 
channel map.  In the following sections we will deal with the 
maser clouds themselves.  
Fig.~\ref{ALL.GPS} shows the
positions of the fitted components in each channel relative to
the estimated stellar position (Section \ref{spos}) for each
source. 
A complete list of the fitted components and their fitted parameters
is available electronically via CDS.

In each source, the majority of components are
resolved. Table~\ref{spotsize} summarises component properties. 
The apparent size of an individual maser component $s$ (measured as described
in Section~\ref{obs+red}) is the FWHM of beamed maser emission from
molecules with velocities within the 0.105~km~s$^{-1}$ channel width,
and is smaller than the physical size of the emitting region. The
greater the maser amplification factor, the smaller the beaming angle
\cite{Elitzur92}, and so under comparable conditions brighter maser
components appear smaller. Column (2) gives the total number $N_{\rm
c}$ of 22-GHz maser components in each source and column (3) gives the
fraction of components $F_{\rm c}$ for which $s > \sigma_{\rm s}$,
where $\sigma_{\rm s}$ is the uncertainty in $s$. The error-weighted
mean and scatter of $s$, $\overline{s}$ and $\sigma_{\overline{\rm s}}$
are given in columns (4) and (5).  The scatter is intrinsic to the
data, not simply due to measurement uncertainties. Note that if RT Vir
is at half the distance of the other sources, $\overline{s}$
represents a similar actual size in all sources apart from
U~Ori, whose components appear significantly larger. \scite{Imai97a} resolved out well over half the flux from RT
Vir using VLBI, but established that the brighter spots detected
(mostly $>100$ Jy in the total power spectrum) had a spatial FWHM of
0.3--1.2~mas, which is consistent with our results.

The brightness temperature $T_{\rm Bc}$ of each component was
calculated from its total flux density $S_{\rm c}$ and its size $s$. The
minimum $T_{\rm Bc}$ measured for each source was $10^5$ K; the maxima
are given in column (6) of Table~\ref{spotsize}. For a spherical
masing region, the relationship between beamed component size and
brightness temperature can be approximated by
\begin{equation}
s\propto \alpha\log T_{\rm Bc}. 
\label{beaming}
\end{equation}
 \scite{Elitzur92} develops a more complete model giving
$-1\la\alpha\la-2$ for spherical clouds, explaining how lower values
correspond to more saturated maser emission if the other relevant
factors are unchanged.

\begin{figure}
\hspace*{-0.4cm}
\rotatebox{270}{
\resizebox{11cm}{!}{\epsfbox{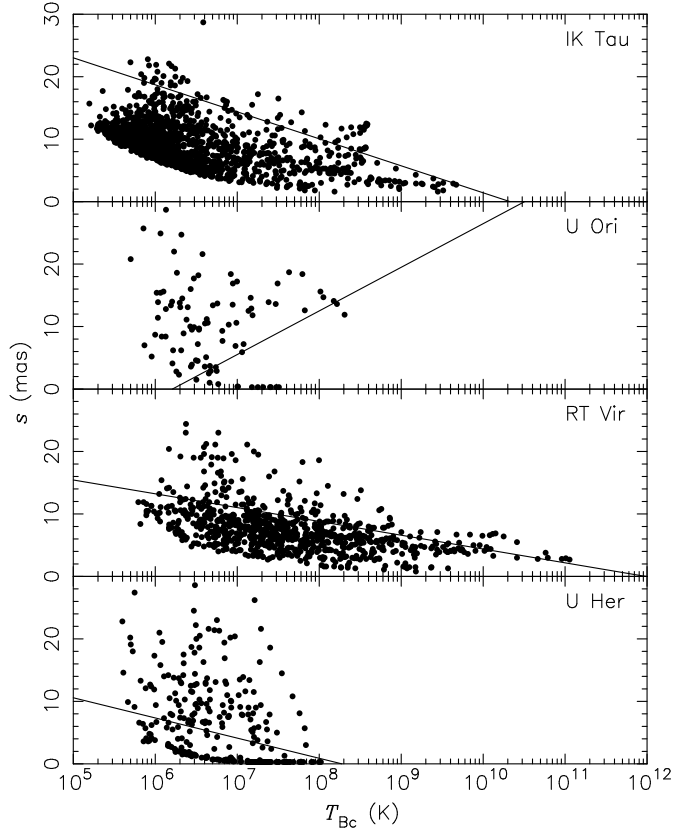}}}
\caption{Maser component FWHM $s$ as a function of brightness
temperature $T_{\rm Bc}$.  The lines show error-weighted fits with
slope $\alpha$ as defined in Equation~\protect{\ref{beaming}}.
}
\label{TBSPOT}
\end{figure}

Fig.~\ref{TBSPOT} shows $s$ as a function of $\log T_{\rm Bc}$.  We
used error-weighted least-squares fits to find the slope $\alpha$,
which is given in column (7) of Table~\ref{spotsize}.  For IK~Tau,
U~Her and RT~Vir, the results are broadly consistent with our
simplified model and imply that the masers are saturated. The formal
errors in $\alpha$ are $\le0.2$ and values of $\alpha<-2$ may be due
to the intrinsic scatter in $s$ and to deviations from our
assumptions. However, for U Ori $\alpha>0$ showing more drastic
revision is needed, and this is discussed in Section~\ref{amp}.

\begin{table}
\caption{The error-weighted statistics of  maser components and
their brightness temperatures, see text for details.}
\begin{tabular}{lllllll}
\hline
Source 	&$N_{\rm c}$ 	&$F_{\rm c}$	& $\overline{s}$ 	& $\sigma_{\rm \overline{s}}$
	&Max. $T_{\rm Bc}$	&$\alpha$ \\
		&	&	&(mas) 	&(mas) 	&(K)  	& \\
 (1) 	&(2) 	& (3) 	&(4)	&(5)	&(6)	&(7)\\
\hline
IK Tau 	&1490	&0.60	& 3 	& 2 	&$5\times10^9$	& $-4$\\
U Ori 	&95 	&0.82	& 9 	& 3 	&$2\times10^8$	& $+7$\\
RT Vir 	&847 	&0.73	& 6 	& 2 	&$1\times10^{11}$	&$-2$\\
U Her 	&282 	&0.55	& 3 	& 2 	&$1\times10^8$	& $-3$\\
\hline
\end{tabular}
\label{spotsize}
\end{table}

\subsection{Measurements of maser clouds}
\label{clouds}

Fig.~\ref{ALL.GPS} shows the grouping of components into 
necklace-like strings, with similar
velocities and positions.   Each string is thought to be produced by 
a single maser cloud, with the emission in each frequency channel 
picking out the line of greatest 
amplification within the velocity range of that channel.  
IK~Tau contains a few clouds with distinct
velocity-position gradients, but most have an 
apparently random distribution of components with velocity.  
In U~Ori, only two
clouds have any clear structure.  However, in U~Her much of the
blue-shifted emission is less random and in RT Vir the majority of
clouds have smooth (not necessarily linear) gradients, which are also
seen in Fig~\ref{RV}.

The measurements of each maser cloud derived from its constituent
components are given in Tables available electronically from CDS. Columns
(1) and (2) give the flux-weighted mean $V_{\rm LSR}$ of the cloud
($\overline{V_{\rm LSR}}$) and its total velocity width
($\Delta{V}$). Columns (3) and (4) respectively give the
FWHM ($\Delta{V_{1/2}}$) of a Gaussian curve fitted to the cloud
velocity profile and the uncertainty in this ($\sigma_{\rm
\Delta{V}_{1/2}}$). Columns (5)--(12) give the position of the
error-weighted centroid of the cloud with respect to the assumed
stellar position (Section~\ref{spos}): $x$, $\sigma_{x}$,
$y$ and $\sigma_{y}$ are the offsets and uncertainties in
Cartesian coordinates and $a$, $\sigma_{\rm a}$, $\theta$ and
$\sigma_{\theta}$ are in polar coordinates. Columns (13) and (14)
respectively give the total angular extent of the cloud ($L$) and its
uncertainty ($\sigma_{\rm L}$); column (15) gives the peak flux
density ($I$), column (16) gives the total flux in the cloud ($S$) and
column (17) gives the peak brightness temperature ($T_{\rm B}$).

Using a distance of 133~pc for RT~Vir and 266~pc for the other
sources, cloud angular size $L$ was converted to the cloud diameter
$l$, which is likely to represent at least 85~per~cent of the true
unbeamed size \cite{Richards99}.
 
\begin{figure}
\vspace*{0.5cm}
\hspace*{-0.8cm}
\rotatebox{-90}{\resizebox{14cm}{!}{\epsfbox{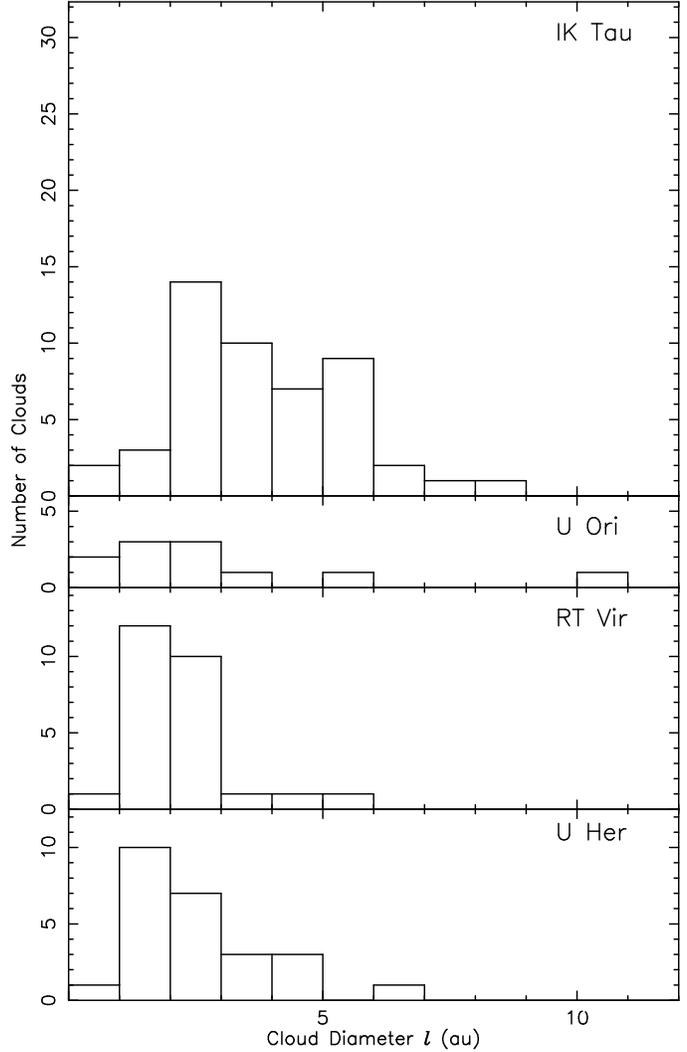}}}
\caption{The distribution of diameters $l$ of maser clouds in each
source, for clouds with $l>1.5\sigma_{\rm l}$.  }
\label{CLOUDL}
\end{figure}

\begin{figure}
\hspace*{-0.5cm}
\rotatebox{-90}{\resizebox{14cm}{!}{\epsfbox{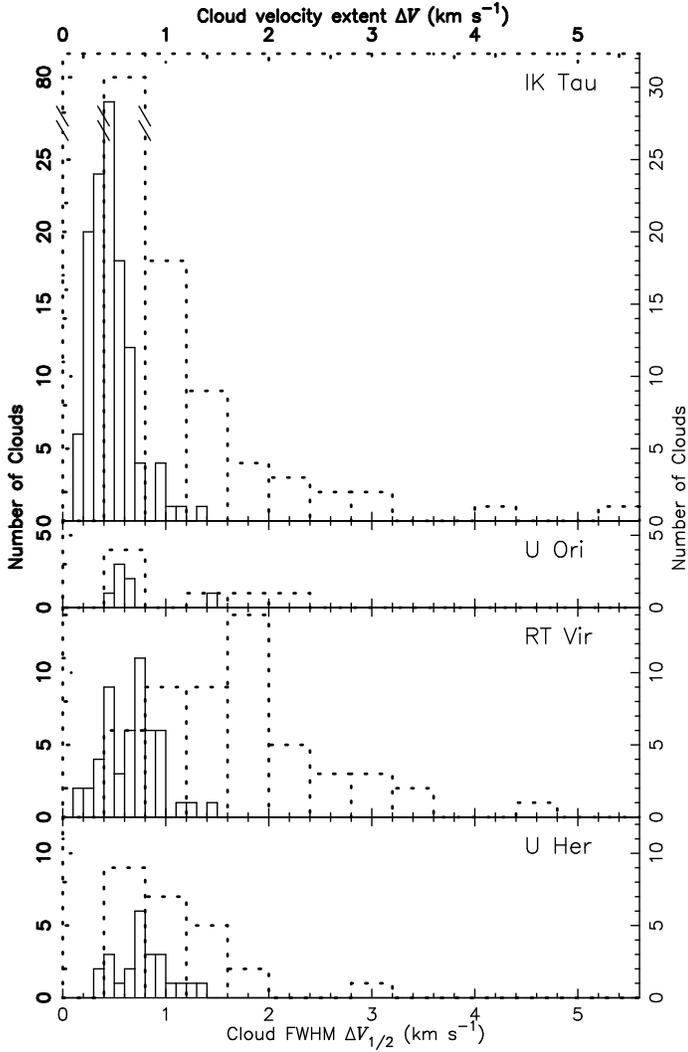}}}
\vspace*{0cm}
\caption{The distribution of the total velocity extent $\Delta{V}$ (dotted lines) and FWHM $\Delta{V_{1/2}}$ (solid lines) of maser clouds in each
source for clouds with $\Delta{V_{1/2}}>1.5\sigma_{\rm
\Delta{V_{1/2}}}$.
}
\label{FWHM}
\end{figure}

Figs.~\ref{CLOUDL} and~\ref{FWHM} show the distribution of $l$,
$\Delta{V_{1/2}}$ and $\Delta{V}$ for each
source. Individual cloud measurements of accuracy $\le1.5\sigma$
(mostly for clouds comprising only 3 components) are not included.
Table~\ref{cloudstats} gives the properties of an average cloud around
each star derived from error-weighted means of the individual cloud
measurements above the threshold of $1.5\sigma$. Column (2) gives the
total number $N$ of clouds in the source. Column (3) gives the number
of clouds $N(l,1.5)$ with $l/\sigma_{\rm l}>1.5$, where $\sigma_{\rm
l}$ is the uncertainty in $l$. The mean ($\overline{l}$) and
dispersion ($\sigma_{\rm \overline{l}}$) of their diameters are given in
columns (4) and (5) respectively. Columns (6) gives the number of
clouds $N(\Delta{V_{1/2}}1.5)$ with $\Delta{V_{1/2}}>1.5\sigma_{\rm
\Delta{V_{1/2}}}$.  These were used to find the mean
($\overline{\Delta{V_{1/2}}}$) and dispersion ($\sigma_{\rm
\overline{\Delta{V_{1/2}}}}$) of the velocity FWHM and of the total velocity
width ($\overline{\Delta{V}}$) and the uncertainty in this
($\sigma_{\rm \overline{\Delta{V}}}$), given in columns (7) -- (10).

Table~\ref{cloudstats} shows that the clouds around IK~Tau have a
significantly larger $\overline{l}$ and a smaller
$\overline{\Delta{V}}$ than those around RT~Vir.  U~Ori and
U~Her have intermediate values.  The IK~Tau clouds also have the
smallest $\overline{\Delta{V_{1/2}}}$.  Fig.~\ref{CLOUDL} shows that
IK~Tau contains a greater number of larger clouds than the other
sources.  Fig.~\ref{FWHM} shows that RT~Vir and U~Her have more clouds
with $\Delta{V_{1/2}}>0.7$~km~s$^{-1}$ than the other sources although
RT~Vir has almost as many with $0.4<\Delta{V_{1/2}}<0.5$~km~s$^{-1}$,
which is the peak range for IK~Tau.  The peak of the distribution of
$\Delta{V}$ is at $1.6-2.0$~km~s$^{-1}$ for RT~Vir, three
times the value for the other sources.  RT~Vir and IK~Tau have a few
clouds with $\Delta{V}\approx3\Delta V_{\rm th}$ where
$\Delta V_{\rm th}$ is the thermal line-width of H$_{2}$O at 1000~K,
$\sim 1.4$~km~s$^{-1}$.


The necklace-like series of components in
Fig.~\ref{ALL.GPS} appear to be in random directions.  
We searched for systematic velocity gradients within maser 
clouds as follows.   For each cloud we attempted to
fit a straight line to $u=V_{\rm LSR}-V_*$ as a function of position
to find the gradient $u'$, taking the goodness of fit
as well as measurement errors into consideration in finding the
uncertainty $\sigma_{\rm u'}$.  Columns (11), (12) and (13) of
Table~\ref{cloudstats} give the number of clouds $N(u')$ with
$u'>2\sigma_{\rm u'}$ used to find the mean ($\overline{u'}$) and the
maximum (Max. $u'$) of $u'$.  
The minimum measurable $\overline{u'}$
was $3\times10^{-12}$~km~s$^{-1}$ m$^{-1}$.  
The typical velocity gradient  $u'$ 
was $\sim$$20\times10^{-12}$~km~s$^{-1}$ m$^{-1}$.  This is at least 
an order of magnitude larger than the systematic velocity 
gradient $K_{\rm v}$ (Table~\ref{rvtab})
within each expanding H$_{2}$O envelope.  
The most 
likely cause of this is local turbulence, which could be thermal, as
$\Delta{V}$ is of the same order as $\Delta{V_{\rm th}}$,   
or due to 
shocks or differential acceleration between regions of different
density or dust:gas ratio (see Section~\ref{quench}).

RT~Vir has the strongest velocity gradients, with 
$\overline{u'}$ more than double the $u'$ seen for the other stars.

\begin{table*}
\caption{The average properties of clouds around each source. See Section~{\protect{\ref{clouds}}} for details. 
}
\begin{tabular}{lrrrrrrrrrrrr}
Source&$N$&$N(\rm l1.5)$ &$\overline{l}$&$\sigma_{\rm \overline{l}}$&$N(\Delta{V_{1/2}}1.5)$&$\overline{\Delta{V_{1/2}}}$& 
$\sigma_{\rm \overline{\Delta{V_{1/2}}}}$&$\overline{\Delta{V}}$&$\sigma_{\rm \overline{\Delta{V}}}$&$N(u')$&$\overline{u'}$&
Max. $u'$\\
&&&\mc{2}{c}{(au)}&&\mc{2}{c}{(km s$^{-1}$)}&\mc{2}{c}{(km
s$^{-1}$)}&&\mc{2}{c}{($10^{-12}$ km s$^{-1}$ m$^{-1}$}) \\
(1)&(2)&(3)&(4)&(5)&(6)&(7)&(8)&(9)&(10)&(11)&(12)&(13)\\
\hline
IK~Tau& 256  & 49 &3.8 & 0.2& 120& 0.46& 0.02   &	0.84 &0.06 &	31&11  & 68 \\	
U~Ori & 14  &  10 &3.0 & 0.8& 7&  0.68 &0.12   &	1.07 &0.24 &	 3&17  & 45\\
RT~Vir &55 &   26 &2.2 & 0.2&  52& 0.67 & 0.04 & 	1.67 &0.11  &	18&43  & 190 \\
U~Her  & 34 &  24 &2.6 & 0.3& 24&  0.78 &0.05   &	1.03 &0.12 &	9&11  & 23 \\
\hline
\end{tabular}
\label{cloudstats}
\end{table*}

We also decomposed $u'$ into d$u$/d$a$ and d$u$/d$\theta$, but found
no statistically significant preferred direction for the individual cloud
velocity gradients.

\begin{figure}
\hspace*{-0.5cm}
\resizebox{9cm}{!}{\epsfbox{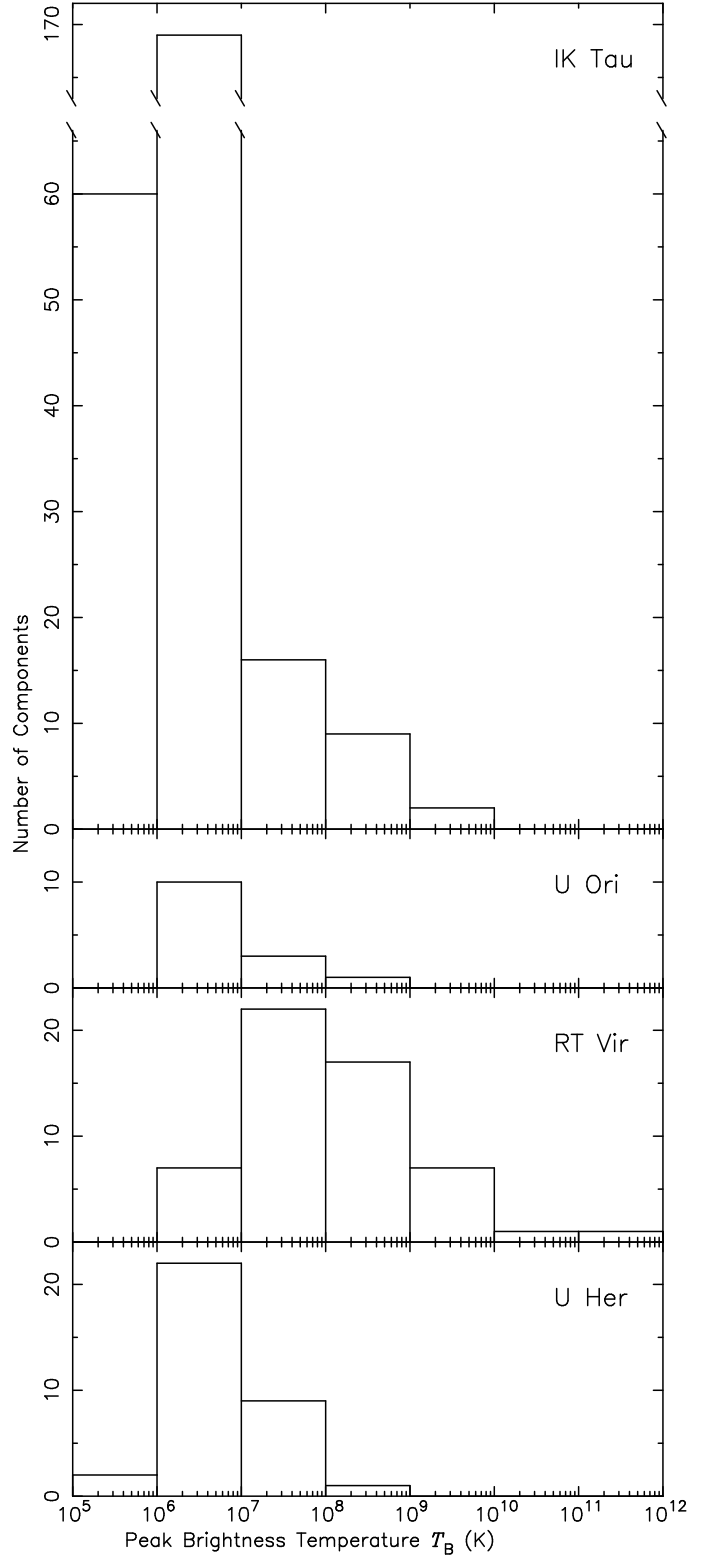}}
\caption{The distribution of the peak brightness temperatures of maser clouds in each source.  }
\label{TBF.PS}
\end{figure}

Fig.~\ref{TBF.PS} shows that the distribution of peak $T_{\rm B}$ for
clouds around RT~Vir reaches a maximum in a range $10\times$ hotter
than for the other stars, as well as extending to values $100\times$
higher. For unsaturated masers, $\Delta{V_{1/2}}$ narrows as the maser
gets brighter and then rebroadens as the maser saturates
(\pcite{Elitzur92}). Although Figs.~\ref{FWHM} and~\ref{TBF.PS}
suggest a correlation between greater $\Delta{V_{1/2}}$ and $T_{\rm
B}$, there is no clear relationship within the cloud measurements for
each source.

\subsection{Cloud densities and survival}
\label{quench}

The inner radius $r_{\rm i}$ of 22-GHz H$_{2}$O maser emission is
determined by the quenching number density $n_{\rm q}$, above which
the collision rate exceeds the radiative decay rate of the upper
level of the maser transition.  At 1200~K, for a fractional H$_2$O
number density $f=4\times10^{-4}n_{\rm H_2}$ (where $n_{\rm H_2}$ is
the H$_2$ number density), $n_{\rm q}=5\times10^{15}$
\cite{Yates97}. \scite{Cooke85} predicted that $r_{\rm
i}\propto\dot{M}^{2/3}/v_{\rm i}$. This relation is plotted in
Fig.~\ref{MDOT}, using the relevant values given in
Tables~\ref{sources} and~\ref{rvtab}. Fig.~\ref{MDOT} shows that the
observed $r_{\rm i}$ of the four sources lie well above the inner
radii inferred from $\dot{M}_{\rm CO}$. This suggests the maser clouds
are denser than average for a homogenous wind, as if they were
produced by localised episodes of mass loss at a higher rate.

The required mass loss rates ($\dot{M}_{\rm H_2O}$, uncertainty
$\sigma_{\rm {\dot{M}_{\rm H2O}}}$) are derived as described in
\scite{Richards99}.  These are given in column (2) and (3) of
Table~\ref{density}.  The CO mass loss rates $\dot{M}_{\rm CO}$ are
given in Table~\ref{sources} and the ratio of the cloud density to the
average density at a given radius is $\dot{M}_{\rm H_2O}/\dot{M}_{\rm
CO}=f_{\rho}$, given in column (4) of Table \ref{density}.  The
quenching density also implies that all maser clouds at a given
temperature have the same density, regardless of the mass loss rate.
$\overline{l}$ is used to find the volume filling factor for the
clouds $f_{\rm V}$ given in column (5) of Table~\ref{density}.

We adopt $r_{\rm p}=r_{\rm i}+(r_{\rm o}-r_{\rm i})/4$ as a typical
radius for the best-filled and brighest part of the shell, where the
true cloud diameter is probably closest to the average measured cloud
size $\overline{l}$.  Following \scite{Richards99}, we estimated the
cloud density at $r_{\rm p}$ by extrapolation from $n_{\rm q}$ and
used this to find the typical mass of a cloud of diameter
$\overline{l}$, $M_{\rm H_2O}$, given in column (6).

The cloud sound crossing time $t_{\rm
cs}$ is given for $\overline{l}$ in column (7).  The time $t_{\rm s}$
taken for a cloud to travel from $r_{\rm i}$ to $r_{\rm o}$ is given
in column (8) and used to find the apparent number of clouds formed
per stellar period $N_{\rm P}$, given in column (9).
Fig.~\ref{DPHIDR.PS} shows the time since the clouds left the star as
a function of radius, where the time at $r_{\rm o}$ is time taken to
reach $r_{\rm i}$ (assuming a constant velocity of $v_{\rm i}$ from
$R_*$ to $r_{\rm i}$) plus $t_{\rm s}$.
IK~Tau appears to form more clouds per period than the other stars
although the value for RT~Vir is uncertain due to its irregular
period. 

With $t_{\rm cs}$ of about 1.5--3~yr, all of the sources have $t_{\rm
cs}\sim1/10t_{\rm s}$ making them vulnerable to turbulent disruption.
Indeed, observations at other epochs (see references in
Sections~\ref{IK} -- \ref{UH}) show that individual clouds can be
identified at intervals of up to 18~months but not after 2--5~yr,
depending on the source. There is only indirect evidence at present as
to whether this indicates that the clouds dissipate, or whether their
maser emission in our direction simply decreases. The small volume of
the shell occupied by clouds ($f_{\rm V}<0.01$), see
Table~\ref{density}, together with the long transit time for gas
through the H$_{2}$O maser zone compared with the sound-crossing time
for a cloud ($t_{\rm s}$ and $t_{\rm cs}$ typically 20 yr and 2 yr
respectively) suggest that a change in the maser output towards us is
the most likely explanation.  Such a change could easily come about
due to the large internal velocity gradients $u'$ that we observe in
random directions within individual maser clouds
(Table~\ref{cloudstats} and Section~\ref{clouds}).  A direct proof of
this hypothesis would be to observe clouds reappear at a later date.

The ratio
$\overline{l}/r_{\rm p}$ is remarkably similar for all four stars,
between 0.12 -- 0.20, where the scatter is as expected from the
measurement uncertainties.  If the clouds originate from the star
their birth radius is a fraction of the stellar radius
given by $\overline{l}(R_*/r_{\rm p})/2R_* = 0.08\pm0.02R_*$.

Returning to Table \ref{density}, the total mass in the clouds we
observe in each CSE is given in column (10) by $M_{\rm tot\; H_2O} =
M_{\rm H_2O}N$ ($N$ is given in Table~\ref{cloudstats}).  The total
mass injected at $r_{\rm i}$ during $t_{\rm s}$ is $M_{\rm tot\; inj}
= t_{\rm s}\dot{M}_{\rm CO}$, given in column (11). The uncertainty in
quantities derived from our measurements are $\la50$~per~cent, and
comparisons with $\dot{M}_{\rm CO}$ may be less accurate.

Even allowing for such uncertainties,
some trends observed in all four CSE are  strongly significant.
\begin{itemize} 
\item{The maser clouds are one or two orders of
magnitude denser than the surrounding gas, showing that in all these
stars, mass is lost in the form of density-bounded discrete clumps.}  
\item{$M_{\rm
tot\; inj}/M_{\rm tot\; H_2O}$ is between 1 and 5 (to within our
accuracy of 50~per~cent), consistent with the filling factor and
overdensity of the H$_{2}$O maser clumps.}  
\item{For each star the radius of the clouds, extrapolated back to the stellar
  surface, is $\sim0.1 R_*$.}
\end{itemize}

Prior to this work, the size of Mira clouds had not been
measured. However, we (e.g. \pcite{Richards99};  Murakawa et
al. 2002, MNRAS, submitted)
have previously shown that RSG H$_{2}$O maser clouds are overdense by
a similar amount, and are a similar relative size ($\sim 10$ per~cent)
when extrapolated back to $R_*$, although the absolute values of $l$,
$R_*$, $r_{\rm i}$ and $\dot{M}$ are an order of magnitude greater for
RSG. However, as befits their larger size, RSG cloud diameters are
greater than the velocity resonance length and they survive for at
least 9 years (Richards Yates \& Cohen 1996, 1998).\nocite{Richards96}
\nocite{Richards98v}

These results suggest that the cloud size is not primarily determined
by local physical or molecular processes in the wind such as shocks or
cooling, as these would operate on the same scales around RSG as
around AGB stars. It could be that clumps somehow grow between the
star and $r_{\rm i}$, so the larger $\dot{M}$ and $r_{\rm q}$ of RSG
means the clouds reach a larger mass.  Alternatively, the cloud size
could be directly determined at the stellar surface, related to
convection cells (possibly producing local chemical enrichment),
pulsation irregularities producing ejections, or cooling above
starspots \cite{Frank95}.

\begin{table*}
\caption{The estimates of mass loss rate from H$_{2}$O maser measurements compared with those using CO observations, the maser cloud filling factor and other properties. See Section~{\protect{\ref{quench}}} for details.}
\begin{tabular}{lrrrrrrrrrr}
\hline
Source & $\dot{M}_{\rm H_2O}$     &$\sigma_{\rm {\dot{M}_{\rm H2O}}}$ &$f_{\rho}$
&$f_{\rm V}$ &$M_{\rm H_2O}$   &$t_{\rm cs}$ &$t_{\rm s}$ &$N_{\rm P}$&$M_{\rm tot\; H_2O}$ &$M_{\rm tot\; inj}$  \\ 
       &\mc{2}{c}{($10^{-6}$M$_{\odot}$ yr$^{-1}$)}   &	     &
&($10^{-9}$ M$_{\odot}$) &(yr)	       &(yr)	    & &($10^{-6}$M$_{\odot}$)&($10^{-6}$M$_{\odot}$)	     \\
(1) &(2)&(3)&(4)&(5)&(6)&(7)&(8)&(9)&(10)&(11)\\
\hline 
IK~Tau(a)&$117$  &$19$       & 50        &0.0043	 &257  &3.0         &26         &11	&66	&68	\\
IK~Tau(b)&$25$   &$10$       & 10        &0.0043	 & 62 &3.0         &29         &10	&16	&75	 \\
U~Ori    &$25$   &$12$       & 110       &0.0014	 & 141 &2.4         &24         &1	&2	&6 	\\
RT~Vir   &$15$   &$2$       & 115       &0.0044	 	&44  &1.7         &14        &2		&2	&2 	\\
U~Her    &$83$   &$14$       & 240       &0.0008	 &104  &2.0         &23         &1	&4	&8 	\\
\hline
\end{tabular}
\label{density}
\end{table*}

\subsection{Maser amplification}
\label{amp}

Maser amplification occurs over a velocity resonance length $\delta l$
where $V_{\rm LSR}$ changes by $\Delta V_{\rm th}$. Inspection of the
individual channel maps (Figs. \ref{IKTAU}--\ref{RTVIR}) shows that
there is only occasional blending, and the low filling factors imply
there is $<1$~per~cent chance of extra amplification due to one cloud
overlapping another along the line of sight unless their distribution
is very non-random. The actual clouds could be any
shape, but for simplicity we assume they are spherical (however see
Section~\ref{amp}). 
  Using the method applied to the RSG S~Per by
\scite{Richards99} and the shell parameters in Table~\ref{rvtab}, we
estimate $\delta l$ for each source at $r_{\rm p}$ in the limiting
cases of a cloud along the line of sight to the star (radial beaming)
and in the plane of the sky with the star (tangential beaming).  We
find for IK~Tau $6\ga\delta l\ga5$, for U~Ori $\delta l\approx6$, for
U~Her $10\ga\delta l\ga5$ and for RT~Vir $4\ga\delta l\ga3$~au.  The
radially-beamed $\delta l$ exceeds the tangentially-beamed value by a
larger amount for smaller values of the logarithmic velocity gradient $\epsilon$.  In all cases $\delta
l>\overline{l}$ (Table~\ref{cloudstats}), so the velocity coherence
depth is not limiting the amplification path, unless there is some
systematic asymmetry in the distribution of cloud shapes.

If the only force on clouds is due to radiation pressure on dust,
their diameter in the tangential and radial directions will increase
proportional to $r/r_{\rm i}$ and $(r/r_{\rm i})^{\epsilon}$
respectively.  If the clouds are on average initially spherical, they
will retain their shape where $\epsilon$ is not much less than 1.
However, the clouds near $r_{\rm i}$ could be flattened by shocks due
to the stellar pulsations. This favours tangential 22-GHz emission,
perpendicular to the direction of the shock, not only due to the
longer path length for maser amplifcation, but also because the pump
IR photons escape along the short axis (\pcite{Elitzur92a}).  In this
geometry, the beaming angle is independent of $T_{\rm B}$ and the
observed component angular size is equivalent to the projected width
of the flattened cloud (\pcite{Elitzur92}).

Thus, shock compression would explain the large scale of emission
inferred from the visibilities in the brightest channel of U~Ori
(Section~\ref{UO}) and its large values of $\alpha$ and
$\overline{s}$ in Table~\ref{spotsize}. The ring-like appearance of
the masers around U~Her is also consistent with shocked clouds.  For
unsaturated masers, $T_{\rm B}$ is roughly exponentially proportional
to a linear function of the path length.  The brightest masers around
U~Her have $T_{\rm B}\approx10^8$~K.  If this arises from clouds with
an axial ratio of 2:1 viewed along the long axis, clouds viewed along
the short axis will have $T_{\rm B}\approx10^4$ K, so clouds lying
close to the line of sight to the star are well below the detection
threshold.  The average cloud diameter $\overline{l}$ was calculated
on the basis of spherical clouds.  If we only detect flattened clouds,
$l$ becomes the long axis, leading to an overestimate of
$\overline{l}$. However, if the clouds around the front and back caps
are present but not detected, then $N$ and $N_{\rm P}$ are
underestimated but the accuracy of other quantities given in
Table~\ref{density} is not greatly affected.

The 22-GHz photon luminosity rate from the whole of each shell,
$\Phi$, is shown in column (13) of Table~\ref{rvtab}. For the Miras,
the average $d\Phi/dr$ is
$(0.3-0.6)\times10^{30}$~photons~s$^{-1}$~m$^{-1}$ but for RT~Vir it
is $3\times10^{30}$~photons~s$^{-1}$~m$^{-1}$, i.e. $\la10\times$
higher. The individual masers around RT~Vir also reach the highest
$T_{\rm B}$, although this object has the lowest $\dot{M}$, the
smallest shell and the smallest clouds.  It also has the smallest,
most irregular, stellar pulsations. It shows the largest velocity
gradients ($u'$ and $K_{\rm v}$) and it is the only source with
$\overline{\Delta{V}}>\Delta{V_{\rm th}}$
(Table~\ref{cloudstats}). The masers are very rapidly variable,
suggesting highly efficient but unsaturated amplification. Small
clouds with a steep velocity gradient could allow the clouds to remain
optically thin at the IR pump photon frequencies (\pcite{Neufeld91},
\pcite{Elitzur92}) and achieve very efficient 22-GHz maser
pumping. This is analogous to the suggested enhanced masing from
shock-compressed clouds in U~Her and U~Ori.  Shock compression alone
is a less likely explanation in the case of RT~Vir as its bright
masers extend well beyond $r_{\rm i}$ and its optical period and
amplitude are poorly defined and small (Table~\ref{sources}),
suggesting the stellar pulsations are either weak or very complex.

Another possiblity is that the wind from RT~Vir is richer in
H$_{2}$O. The latter could be due to a higher O:C ratio, suggesting
RT~Vir had a higher progenitor mass and/or is more highly evolved than
the Miras (since all 4 stars are so nearby that the overall Galactic
metallicity gradient has little effect although local fluctuations are
possible). \scite{Cooke85} showed that, for unsaturated masers,
$d\Phi/dr$ increases by up to an order of magnitude if the fractional
abundance of H$_2$O, $f$=[H$_2$O]/[H$_2$], increases by a factor of 2.
$n_{\rm q}$ is weakly dependent on $f$, so the clouds around RT~Vir
could be even denser by a factor of a few tenths. This could increase
the acceleration of the gas, but only slightly, as the momentum
coupling with the dust is already good for most of the density range
in the H$_{2}$O maser clouds (\pcite{Richards99}).

For IK~Tau and RT~Vir we were able to estimate the 
photon luminosity as a function of distance from the star 
(Section~\ref{limits}).  This is shown in Fig.~\ref{DPHIDR.PS}.  
Comparing our results with the models by  \scite{Cooke85} we found 
that the emissivity was 100-1000 times greater than predicted for the 
mass-loss rates based on CO, $\dot{M}_{\rm CO}$.
This difference can be explained in terms of the clumping of the 
mass-loss into the discrete clouds we observe.  
The collision rate
(and hence the maser pump rate) is proportional to $\rho^2$, where
$\rho$ is the local density.  Thus, the postulated over-density of the maser
clouds should increase the maser luminosity at the same radius.

From Table~\ref{density}, the IK~Tau maser clouds are over-dense
by a factor $f_{\rho}\sim10-50$, suggesting that the emission at an
equivalent distance from the star should be $\ge100$ times brighter
than the value deduced from $\dot{M}_{\rm CO}$.  This is consistent
with the models of \scite{Cooke85} for a fractional H$_{2}$O density
of $f\le2\times10^{-4}$.  However, other factors may be involved as
the profile of $d\Phi/d{r}$ as a function of $r$ is very irregular,
which implies changes in the rate or composition of mass loss from
IK~Tau as well as possible axisymmetry of the CSE.

For RT~Vir, the clouds are over-dense by a factor of  $f_{\rho}\sim115$.
This suggests that $d\Phi/d{r}$ should exceed the value deduced from
$\dot{M}_{\rm CO}$ by a factor of roughly $115^2/(3^2)^2$,
i.e. 1--200.  In fact, RT~Vir is brighter than the models of
\scite{Cooke85} by 3 orders of magnitude, even for the highest
fractional   H$_{2}$O density considered, $f=6\times10^{-4}$.

\section{Summary and conclusions}
\label{conclusions}

MERLIN 22-GHz observations of the 4 low-mass AGB stars IK~Tau, U~Ori,
RT Vir and U~Her show that the H$_{2}$O masers surrounding each star
are found in thick expanding shells of a few hundred mas in diameter.  The
inner and outer limits to the maser distributions were derived, under
the assumption of spherically symmetric expansion. The inner radii
$r_{\rm i}$ are at 6--16~au, with outer limits $\sim$4 times greater.
The maser velocity increases two- or three-fold over this distance,
giving a logarithmic velocity gradient $\epsilon$ in the range
$0.5\le\epsilon\le1$.

For the first time, the individual 22-GHz masers in each CSE have been
resolved. The unbeamed H$_{2}$O vapour cloud diameter $l$ is about
2--4~au at a distance of 10--25~au from the star. If the clouds are
formed close to the stellar surface, assuming linear expansion, this
corresponds to an original cloud radius of $\sim0.1R_*$.  We find that
$l$ is less than the velocity resonance length for spherical
clouds. The average cloud velocity span is 0.8--1.7~km~s$^{-1}$, with
a FWHM between 0.5--0.8~km~s$^{-1}$.

The water vapour cloud density at $r_{\rm i}$ is one or two orders of
magnitude greater than the mean density derived from the mass loss
rates inferred from CO and other observations. The volume filling
factor of the shells is $<0.5$ per~cent. In addition, we find that the
clouds are vulnerable to disruption by shocks during the passage
through the shell, as the sound crossing time ($t_{\rm cs}$) of the
clouds and the time the cloud takes to travel through the maser shell
($t_{\rm s}$) are related by $t_{\rm cs}\sim(1/10)t_{\rm s}$. These
facts suggest that the masers are density bounded. We find that the
maser survival times and sizes of clouds around these AGB stars
are $\sim 10$ times smaller than those around RSG, approximately the
same ratio as the stellar masses and sizes. This implies that it is
the properties of the central star that determine maser cloud sizes,
rather than behaviour intrinsic to the wind.

U~Ori and U~Her have poorly-filled H$_{2}$O maser shells and both have
shown OH maser flares. Some individual 22-GHz maser components in
U~Ori have a large beamed size which does not decrease with increasing
$T_{\rm B}$, indicative of masing from a slab rather than a sphere. In
U~Her, the masers form a hollow ring. The appearance of the 22-GHz
masers in both of these sources suggests that they emanate from
shock-compressed shells, producing strong tangential amplification;
such shocks could be the cause of the OH flares.

IK~Tau and RT~Vir have well-filled H$_2$O maser shells. Observations
of both of these sources made $\la$ 10 years apart show a persistent
E-W offset between moderately red- and blue-shifted emission, although
individual masers do not survive more than $\sim1.5$ yr. This can be
explained by an equatorial density enhancement in a spherical shell,
so that the brightest masers lie in an oblate spheroid and the
equatorial plane is at an angle $i$ to the line of sight, such that
$\pi/2 > i > \pi/4$. At any one epoch, the shells of U~Her and U~Ori
appear asymmetric, but after comparing our maps with those published
at different epochs (see Section~\ref{discuss} for references), we
cannot distinguish any systematic deviations from poorly-filled
spherical shells.
 
We are obtaining multi-epoch maps of H$_{2}$O and OH masers using
MERLIN and the EVN to provide similar resolution for all species. This
will allow us to study the evolution of individual maser clouds.  We
will measure their proper motions and distinguish between acceleration
and any latitude dependence of expansion velocity e.g. a bipolar
outflow. We will thus examine the stability of the velocity field
  in the CSE as a whole and maser amplification within individual clouds.
We will measure changes in maser size and brightness with
time and so constrain the degree of maser saturation and the local
overdensity of maser clouds as a function of radius.  We will develop
a fuller model of the velocity fields to compare with predictions
based on dust properties and distribution.  

\section{Acknowledgements}

MERLIN is the Multi Element Radio Linked Interfermometer Network, a
national facility operated by the University of Manchester at Jodrell
Bank Observatory on behalf of PPARC.  We thank the MERLIN staff for
performing the observations. In this research we have used, and
acknowledge with thanks, data from the AAVSO database submitted to the
AAVSO by variable star observers worldwide.  DRG gratefully
acknowledges a research grant from CONACYT, the Mexican Research
Council, and an EC Marie Curie studentship at the IoA Cambridge.

\bibliography{cse}
\bsp
\label{lastpage}
\end{document}